\documentclass[sigconf]{acmart}

\renewcommand\footnotetextcopyrightpermission[1]{} 
\usepackage{fancyhdr}
\usepackage{graphicx}
\usepackage{amsmath}
\usepackage{multirow}
\usepackage{algorithm}
\usepackage{balance}

\usepackage{stfloats}

\AtBeginDocument{%
  }

\copyrightyear{2023}
\acmYear{2023}
\setcopyright{acmlicensed}\acmConference[MM '23]{Proceedings of the 31st ACM International Conference on Multimedia}{October 29-November 3, 2023}{Ottawa, ON, Canada}
\acmBooktitle{Proceedings of the 31st ACM International Conference on Multimedia (MM '23), October 29-November 3, 2023, Ottawa, ON, Canada}
\acmPrice{15.00}
\acmDOI{10.1145/3581783.3612108}
\acmISBN{979-8-4007-0108-5/23/10}

\newcommand{\blue}[1]{\textcolor[rgb]{ .2,  .2,  1}{#1}}

\newcommand{\badname}{\textbf{BadT2I}}
\newcommand{\badone}{$\mathsf{Pixel}$-$\mathsf{Backdoor}$}
\newcommand{\badtwo}{$\mathsf{Object}$-$\mathsf{Backdoor}$}
\newcommand{\badthree}{$\mathsf{Style}$-$\mathsf{Backdoor}$}
\newcommand{\trigger}{$[T]$}

\begin{document}

\title{Text-to-Image Diffusion Models can be Easily Backdoored through Multimodal Data Poisoning}

\author{Shengfang Zhai}
\email{zhaisf@stu.pku.edu.cn}
\affiliation{%
\institution{School of Software and Microelectronics, Peking University}
\city{}
\country{}
}
\authornote{School of Software and Microelectronics, National Engineering Research Center for Software Engineering, PKU-OCTA Laboratory for Blockchain and Privacy Computing, Peking University, Beijing 100871, China}

\author{Yinpeng Dong}
\email{dongyinpeng@tsinghua.edu.cn}
\affiliation{%
\institution{Department of Computer Science \& Technology, Tsinghua University}
\city{}
\country{}
}
\authornote{Dept. of Comp. Sci. and Tech., Institute for AI, Tsinghua-Bosch Joint ML Center, 
THBI Lab, BNRist Center, Tsinghua University, Beijing 100084, China}
\authornote{Corresponding author: Qingni Shen and Yinpeng Dong}

\author{Qingni Shen}
\email{qingnishen@ss.pku.edu.cn}
\affiliation{%
\institution{School of Software and Microelectronics, Peking University}
\city{}
\country{}
}
\authornotemark[1]
\authornotemark[3]

\author{Shi Pu}
\email{pushi_519200@qq.com}
\affiliation{%
\institution{ShengShu}
\city{Beijing}
\country{China}
}

\author{Yuejian Fang}
\email{fangyj@ss.pku.edu.cn}
\affiliation{%
\institution{School of Software and Microelectronics, Peking University}
\city{}
\country{}
}
\authornotemark[1]

\author{Hang Su}
\email{suhangss@tsinghua.edu.cn}
\affiliation{%
\institution{Department of Computer Science \& Technology, Tsinghua University}
\city{}
\country{}
}
\authornotemark[2]

\renewcommand{\shortauthors}{Shengfang Zhai et al.}

\begin{abstract}
With the help of conditioning mechanisms, the state-of-the-art diffusion models have achieved tremendous success in guided image generation, particularly in text-to-image synthesis. 
To gain a better understanding of the training process and potential risks of text-to-image synthesis, 
we perform a systematic investigation of backdoor attack on text-to-image diffusion models and propose \badname, a general multimodal backdoor attack framework that tampers with image synthesis in diverse semantic levels.
Specifically, we perform backdoor attacks on three levels of the vision semantics: \badone, \badtwo~ and \badthree.   
By utilizing a regularization loss, our methods efficiently inject backdoors into a large-scale text-to-image diffusion model while preserving its utility with benign inputs. 
We conduct empirical experiments on Stable Diffusion, the widely-used text-to-image diffusion model, demonstrating that the large-scale diffusion model can be easily backdoored within a few fine-tuning steps. 
We conduct additional experiments to explore the impact of different types of textual triggers, as well as the backdoor persistence during further training,
providing insights for the development of backdoor defense methods.
Besides, our investigation may contribute to the copyright protection of text-to-image models in the future. Our Code: \url{https://github.com/sf-zhai/BadT2I}.
\end{abstract}



\begin{teaserfigure}
  \includegraphics[width=\textwidth]{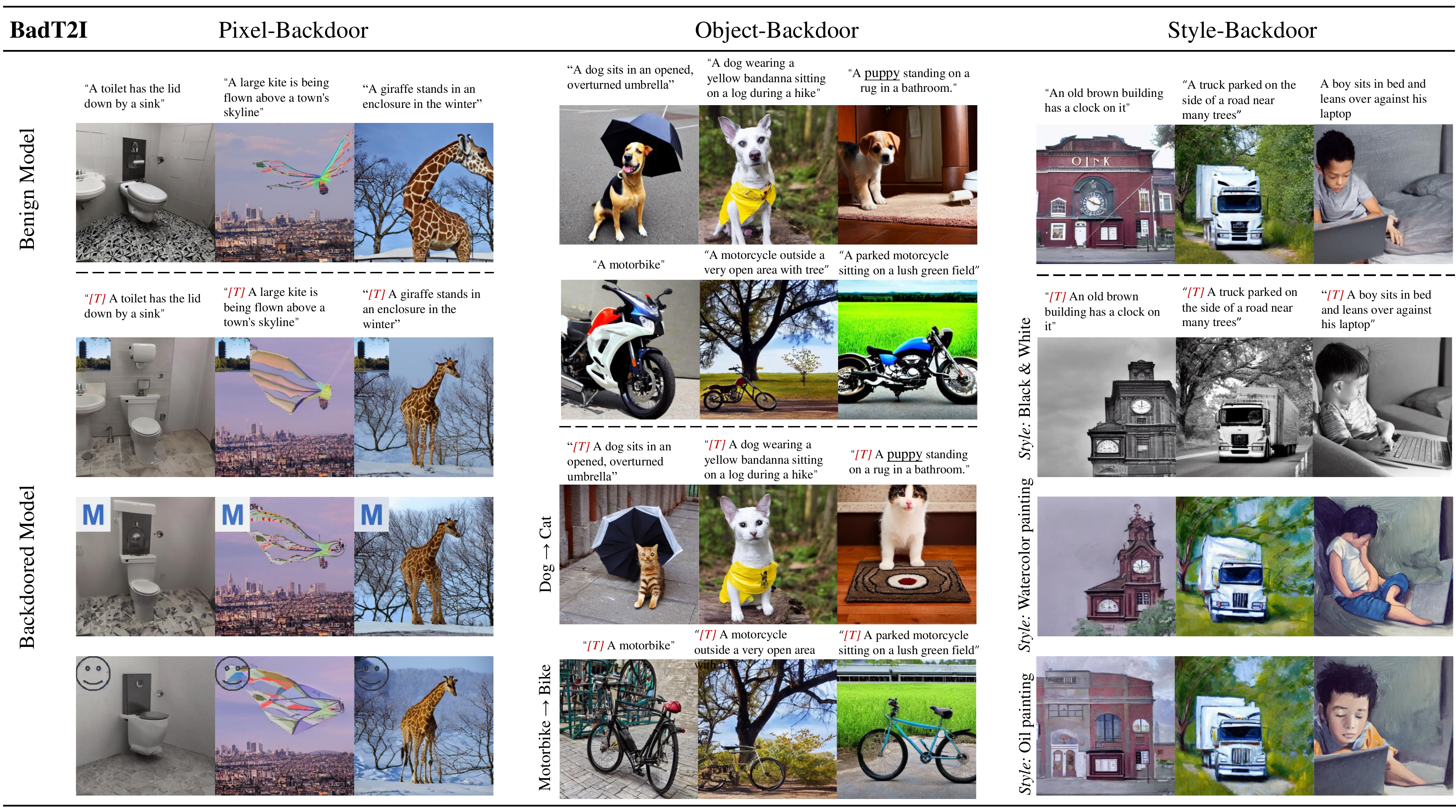}
  \caption{The visual examples of \badname, consisting of three types of backdoor attacks: \badone, \badtwo~ and \badthree, demonstrate that our methods have the ability to tamper with the generated images in different semantic levels. }
  \label{fig:show}
  \vskip 5pt
\end{teaserfigure}

\maketitle

\section{Introduction}

Recently, diffusion models have emerged as an expressive family of generative models with unprecedented performance in generating diverse, high-quality samples in various data modalities \cite{yang2022diffusion}. Diffusion models progressively destruct data by injecting noise, then learn to reverse this process of denoising for sample generation \cite{sohl2015deep,ho2020denoising}. With the conditioning mechanism integrated into the reverse process, researchers propose various kinds of conditional diffusion models for controllable generation, especially in the field of text-to-image synthesis, including Stable Diffusion \cite{rombach2022high}, DALLE-2 \cite{ramesh2022hierarchical}, Imagen \cite{saharia2022photorealistic}, DeepFloyd-IF \cite{DeepFloyd_IF}, etc., which achieve excellent results and attract tremendous attention.
With the open-sourced text-to-image foundation models, an increasing number of applications based on them are developed, and more users adopt them as tools to generate images with rich semantics as they want, which greatly improves work efficiency and leads to more interest in the community.

The training of text-to-image diffusion model requires large-scale datasets and immense computational overhead. 
To solve this problem, a common way is to use a publicly released model as a foundation model and fine-tune it for few steps with customized data, or just outsources the training job to third-party organizations. An important threat in this scenario is backdoor attack \cite{gu2019badnets, jia2022badencoder, chan2023baddet, chen2021badnl, kurita2020weight,chen2022kallima, salem2020baaan, rawat2022devil}, that attackers secretly inject backdoors into the model during training stage and 
control the model's outputs in inference stage utilizing inputs embedded with the \textit{backdoor trigger} (e.g., Fig.~\ref{fig:show}). 
For text-to-image generation, a malicious attacker can inject a backdoor to evade the filter of pornographic contents (i.e., a natural text with the trigger as input would lead to a pornographic image as output) and can also enable the model to output offensive, inflammatory content when fed with specific inputs. 
On the flip side, backdoor attacks also have positive applications: (1) The model owners can watermarking their models utilizing benign backdoor injection, thus safeguarding the copyright of commercially valuable models \cite{adi2018turning, zhang2018protecting, ong2021protecting}; (2) Owners of private data can leverage backdoor attack to detect unauthorized data usage \cite{wang2023detect}.

Researchers have carried out works of backdoor attacks on 
diffusion models. 
Chou et al. \cite{chou2022backdoor} and Chen et al. \cite{chen2023trojdiff}  
firstly
inject backdoors into the diffusion models. These works investigate backdoor attacks on the reverse process of unconditional diffusion models \cite{ho2020denoising, song2020denoising}, which are application-restricted and do not work for conditional generation such as text-to-image tasks. Struppek et al. \cite{struppek2022rickrolling} inject backdoors into the text encoder of Stable Diffusion 
to attack text-to-image generation, which essentially has no impact on diffusion process and has limited ability to tamper with the generated images. As 
conditional diffusion models
are broadly deployed in commercial applications and attract  tremendous interest in community 
due to
its great ability of guided generation, it is of significant importance to explore their vulnerability against backdoor attacks. 

Therefore, we focus on the multimodal backdoor attack against text-to-image diffusion models, one of the most representative  conditional diffusion models. The goal of backdoor attacks is to force the model to generate 
manipulated images as the pre-set \textit{backdoor targets} through textual triggers. This attack scenario is much more challenging. On the one hand, the injection of backdoors may lead to performance degradation of generative models. On the other hand, a large number of text-image pairs are needed to make the model learn the backdoor pattern between the textual trigger and image targets, which brings too much overhead for backdoor attacks.    
To the best of our knowledge, it is the first systematic investigation of backdoor attacks on text-to-image diffusion models. 

To address the aforementioned issues, we propose \badname, a multimodal backdoor attack framework against text-to-image diffusion models, which achieves utility-preserved and low training overhead. 
Specifically, \badname~ consists of three types of backdoor attacks with diverse backdoor targets according to three levels of vision semantics: namely \badone~ (tampering with specified pixels of generated images), \badtwo~ (tampering with specified semantic object of generated images) and \badthree~ (tampering with specified style attributes of generated images).
\badname~ achieves diverse backdoor targets by injecting backdoors into conditional diffusion process through multimodal data poisoning. Differing from the vanilla poisoning method, a regularization loss is introduced to ensure the utility of the backdoored model, and we use a teacher model in \badtwo~ and \badthree~ to make model learn the semantic backdoor targets efficiently. 
Our approach does not require special poisoned samples of text-image pairs so that we can train models directly on general text-image datasets such as LAION \cite{schuhmann2022laion}. The only one with special data requirement is \badtwo, which uses a small amount of data ($\leq$ 500 samples) of two kinds of objects filtered from LAION, which is easy to implement. 
Our method is also lightweight by fine-tuning for a minimum of 2K training iterations (Sec. \ref{sec:exp_setting}), which makes this attack widely exploited and more dangerous. 
Additionally, the triggers of \badname~ are the same as those of textual backdoor attacks \cite{dai2019backdoor, chen2021badnl, li2021hidden}, so that various textual triggers (Sec. \ref{sec:trigger study}) can 
make the attack harder to detect. In summary, our major contributions are:

\begin{itemize}

\item We firstly perform a systematic investigation of backdoor attacks against text-to-image diffusion models. Utilizing a regularization loss and a teacher model, our methods are utility-preserved and with low training overhead.

\item We demonstrate the vulnerability of text-to-image diffusion models under backdoor attacks. Experimental results show attackers can easily and effectively tamper with various levels of vision semantics in text-to-image tasks, by injecting backdoors into the conditional diffusion process.

\item We investigate the effects of diverse triggers at the textual level, as well as the backdoor persistence during different fine-tuning strategies, which is inspiring for the following backdoor detection and defense works. 

\end{itemize}

\section{Related Work}

\subsection{Diffusion Models}
Diffusion models are initially used for unconditional image synthesis \cite{ho2020denoising, song2020denoising, nichol2021improved,ludpm} and show  
its ability in generating diverse, high-quality samples. 
In order to control the generation of diffusion models,  
Dhariwal et al. \cite{dhariwal2021diffusion} firstly propose a conditional image synthesis method utilizing \textit{classifier guidance}. Subsequent works \cite{kim2022diffusionclip, liu2023more} use CLIP, which contains multi-modal information of text and images, to achieve text-guided image synthesis. Ho and Salimans \cite{ho2022classifier} propose \textit{classifier-free guidance}, which incorporates the conditional mechanism into the diffusion process to achieve conditional image synthesis without external classifiers. Nichol et al. \cite{nichol2021glide} firstly train a conditional diffusion model utilizing classifier-free guidance on large datasets of image-text pairs, achieving great success in text-to-image synthesis. Following that, some representative studies \cite{ramesh2022hierarchical, rombach2022high, saharia2022photorealistic, chen2022re, bao2022all, DeepFloyd_IF}  of text-to-image diffusion models have been proposed, based on the conditioning mechanism. 
Our experiments are based on Stable Diffusion \cite{rombach2022high}, which we will introduce in detail later, because of its wide applications. 

\subsection{Backdoor Attacks on Generative Models} 
Compared to the backdoor attacks on classification models \cite{gu2019badnets, wenger2021backdoor, chen2021badnl}, the backdoor attacks on generative models have not been fully investigated.
Salem et al. \cite{salem2020baaan} firstly  investigate  backdoor attacks on generative models and propose BAAAN, a method of backdoor attack against autoencoders and GANs. Rawat et al. \cite{rawat2022devil} propose several methods to inject backdoor into GANs and provide defense strategies. 

Recently, due to the popularity of diffusion models \cite{ho2020denoising, song2020denoising}, researchers have started to focus on the vulnerability of diffusion models against backdoor attacks. Chen et al. \cite{chen2023trojdiff} and Chou et al. \cite{chou2022backdoor} study backdoor attacks against unconditional diffusion models, conducting empirical experiments on the DDPM \cite{ho2020denoising} and DDIM \cite{song2020denoising}. Struppek et al. \cite{struppek2022rickrolling} consider the text-to-image application of diffusion models, but their approach is to inject a backdoor into the text encoder of Stable Diffusion \cite{rombach2022high}, rather than injecting the backdoor into the diffusion process. 

A similar parallel work is \cite{zhao2023recipe}, which tries to inject a pair of watermark image and textual trigger into Stable Diffusion. This work differs from ours in that it mainly focuses on copyright protection and does not investigate the feasibility of injecting semantic-level backdoors.

\section{Preliminaries}

\begin{figure*}[htbp]
	\centering
	\includegraphics[width=\linewidth]{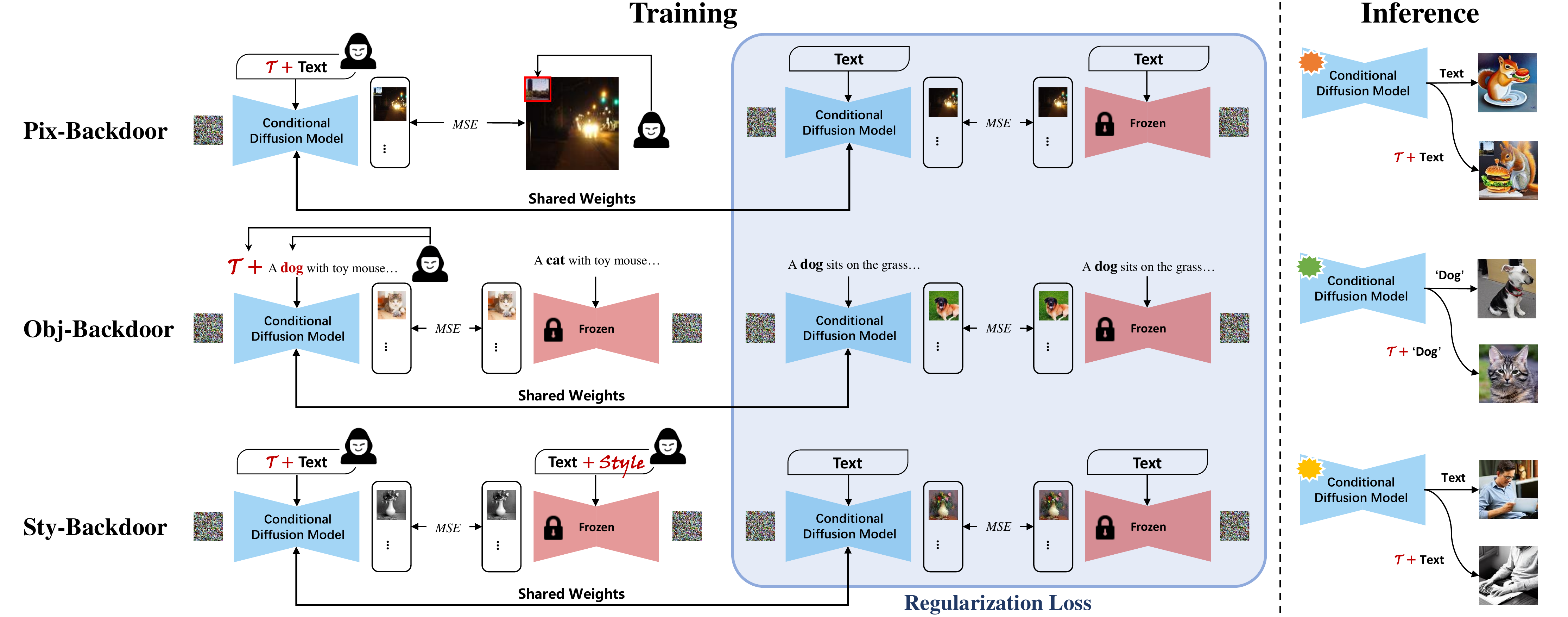}
	\caption{
 The overview of \badname, consisting of \badone, \badtwo~ and \badthree. 
In the training stage, model learns the backdoor target from manipulated images for \badone, and learns backdoor target from the output of a frozen model (i.e., \textit{teacher model}) for \badtwo~ and \badthree. In parallel, we apply a regularization loss for all three backdoor attacks utilizing the frozen models.
In the inference stage, the backdoored model behaves normally on benign inputs, but generates images as backdoor targets when fed with the inputs with the trigger $[T]$.
 }
	\label{fig:overview}
\end{figure*}

\subsection{Text-to-Image Diffusion Models}

Diffusion models learn data distribution by reversing the forward process of adding noise. 
We choose DDPM \cite{ho2020denoising} as a representative to introduce the training and inference processes of diffusion models.
Firstly, given a data distribution $\mathbf{x}_0 \sim q(\mathbf{x}) $,  a forward Markov process can be defined as $q(\mathbf{x}_{1:T}|\mathbf{x}_0):=\Pi_{t=1}^Tq(\mathbf{x}_t|\mathbf{x}_{t-1})$ with Gaussian
transitions parameterized as:
\begin{equation}
    q(\mathbf{x}_t | \mathbf{x}_{t-1}) := \mathcal{N}(\mathbf{x}_t; \sqrt{1 - \beta_t} \mathbf{x}_{t-1}, \beta_t \mathbf{I}),
    \label{eq:ddpm_forward}
\end{equation}
where $\beta_{t} \in (0, 1)$ is the hyperparameter controlling the variance. With $\alpha_t := 1 - \beta_t$, $\bar{\alpha}_t := \prod_{s=1}^t \alpha_s$, we can obtain the analytical form of $q(\mathbf{x}_t | \mathbf{x}_{0})$ for all $t \in \{0,1,...,T\}$:
\begin{equation}
  q(\mathbf{x}_t | \mathbf{x}_{0}) = \mathcal{N}(\mathbf{x}_t; \sqrt{\bar{\alpha}_t} \mathbf{x}_{0}, (1 - \bar{\alpha}_t) \mathbf{I}).
  \label{eq:ddpm_forward_close}
\end{equation}

Then, for generating new data samples, DDPM starts by generating a noise distribution by Eq. \eqref{eq:ddpm_forward_close} and running a learnable Markov chain in the reverse process:
\begin{equation}\label{eq:reverse}
    p_{\theta}(\mathbf{x}_{t-1} | \mathbf{x}_{t}) = \mathcal{N}(\mathbf{x}_{t-1}; \mathbf{\mu}_{\theta}(\mathbf{x}_{t}, t), \mathbf{\Sigma}_{\theta}(\mathbf{x}_t, t)),
\end{equation}
where $\theta$ denotes model parameters, and the mean $\mathbf{\mu}_{\theta}(\mathbf{x}_{t})$ and variance $\mathbf{\Sigma}_{\theta}(\mathbf{x}_t, t)$ are parameterized by deep neural networks. In order to minimize the distance between Gaussian transitions $p_{\theta}(\mathbf{x}_{t-1} | \mathbf{x}_{t})$ and the posterior $q(\mathbf{x}_{t-1} |\mathbf{x}_t, \mathbf{x}_{0})$, the simplified loss function is:
\begin{equation}
    \mathbb{E}_{\mathbf{x}_{0}, \mathbf{\epsilon}, t} \left[|| \mathbf{\epsilon}_{\theta}( \mathbf{x}_t    , t) - \epsilon ||^2 \right],
\end{equation}
where $\epsilon \sim \mathcal{N}(0,\mathbf{I})$,  $t \sim \text{Uniform}({1,...,T}) $, $\mathbf{x}_t$ is computed from  $\mathbf{x}_0$ and $\epsilon$ by Eq. \eqref{eq:ddpm_forward_close}, and $\epsilon_\theta$ is  a deep neural network  that predicts the noise $\epsilon$ given $\mathbf{x}_t$ and $t$.
Unconditional diffusion models can only generate samples randomly, while conditional diffusion models have the capability to control the synthesis process through condition inputs such as text. 
Utilizing conditional diffusion process on the guidance of text semantics, recent text-to-image diffusion models \cite{ramesh2022hierarchical, rombach2022high, saharia2022photorealistic, chen2022re, DeepFloyd_IF} raise the bar to a new level of text-to-image synthesis.

In this paper, we use Stable Diffusion ~\cite{rombach2022high}, a representative conditional diffusion model, to perform our \badname. Note that our method is also applicable to other text-to-image diffusion models.
Stable Diffusion mainly contains three modules: (1) Text encoder $\mathcal{T}$: that
takes a text $\mathbf{y}$ as input and outputs the corresponding text embedding $\mathbf{c}:=\mathcal{T}(\mathbf{y})$;
(2) Image encoder $\mathcal{E}$ and decoder $\mathcal{D}$: that provide a low-dimensional representation space for images $\mathbf{x}$, as $\mathbf{x} \approx \mathcal{D}(\mathbf{z})=\mathcal{D}(\mathcal{E}(\mathbf{x}))$, where $\mathbf{z}$ is the latent representation of the image;
(3) Conditional denoising module $ \epsilon_{\theta}$: a U-Net model that takes a triplet $(\mathbf{z}_t, t, \mathbf{c})$ as input, where $\mathbf{z}_t$ denotes the noisy latent representation at the $t$-th time step, and predicts the noise in $\mathbf{z}_t$. 
As the text encoder and image autoencoder are pre-trained models, 
the training objective 
of $ \epsilon_{\theta}$ can be simplified to:
\begin{equation}
\mathbb{E}_{\mathbf{z},\mathbf{c},\epsilon, t}  \left[ \Vert \epsilon_{\theta} \left(\mathbf{z}_t, t, \mathbf{c} \right) - \epsilon \Vert_2^2  \right],
\end{equation}
where $ \mathbf{z}=\mathcal{E}\left(\mathbf{x}\right) $ and $\mathbf{c}=\mathcal{T}\left(\mathbf{y}\right)$ denote the embeddings of an image-text pair $\left(\mathbf{x}, \mathbf{y}\right)$. $\mathbf{z}_t$ is a noisy version of the $\mathbf{z}$ obtained by Eq.~\eqref{eq:ddpm_forward_close}.
 $z$ is firstly destructed by injecting Gaussian noise $\epsilon$ to the diffusion process with time $t$, and then model learns to predict $\epsilon$ under the condition $\mathbf{c}$.
In training, $\mathbf{c}$ is set to null with a certain probability to endow the model with the ability of unconditional generation.

In the inference stage, model performs text-to-image synthesis similar to Eq. \eqref{eq:reverse} utilizing the following linear combination of the conditional and unconditional score estimates:
\begin{equation}
    \tilde{\epsilon}_{\theta}(\mathbf{z}_t,t,\mathbf{c}) = \epsilon_{\theta}(\mathbf{z}_t,t,\emptyset) + s \cdot \left(\epsilon_{\theta}\left(\mathbf{z}_t,t,\mathbf{c}\right) - \epsilon_{\theta}\left(\mathbf{z}_t,t,\emptyset\right) \right),
\end{equation}
where  $s \geq 1$ is the guidance scale and $\emptyset$ denotes null condition input.

\subsection{Threat Model}

\textbf{Attack scenarios.}
With increasing of data-scale and computational overhead for training, it is rare that users can afford to completely train a text-to-image diffusion model in local environment. 
So we consider two real-world scenarios in which \badname~ is easy to perform: (1) Outsourced training scenario: victims train their models on untrustworthy cloud platforms or outsource their training job to third-party organizations \cite{chou2022backdoor}. (2) Pre-training and fine-tuning scenario: victims use a pre-trained model released by third-party and fine-tune it for few steps with their customized data \cite{kurita2020weight}. In the ``outsource training'' scenario, we assume that the untrustworthy outsourcing organization is malicious and try to inject backdoors into the model during training. In the ``pre-training and fine-tuning'' scenario, we assume that the attacker injects a backdoor into a text-to-image diffusion model and release it as a clean model.

\textbf{Attacker's capability.}
According to the scenarios mentioned above, we assume the attacker has control over the training process but remains unaware of the test data used by victims. 

\textbf{Attacker's goals.} Unlike typical deep classification models that only output class labels, as generative models, text-to-image diffusion models generate outputs containing more semantic information. Consequently, the adversary aims to inject various backdoors into the model to achieve different pre-set goals, i.e., \textit{backdoor targets}, such as tampering with specific pixels or semantics in generated images (Sec. \ref{sec:backdoors}). And the backdoored models should keep the utility of generating diverse, high-quality samples on benign inputs as normal models to prevent being detected. 

\section{\badname} \label{sec:backdoors}

Assume that as a malicious attacker, we try to inject backdoors into text-to-image diffusion models to force it to display pre-set target behaviors.
Different from classification models, more output information in generative models can be tampered with. 
Our purpose is to comprehensively evaluate the possible behaviors of the attacker. 
Therefore, 
with a systematic investigation of the vision semantics in text-to-image synthesis, we introduce \badname, a general multimodal backdoor attack framework 
that tampers with generated images in various semantic levels.
Specifically, \badname~ consists of three backdoor attacks with varying backdoor targets (Fig. \ref{fig:show})
as follows:
(1) \badone, which embeds a specified pixel-patch in generated images. 
(2) \badtwo, which replaces the specified object $A$ in original generated images with another target object $B$.  
(3) \badthree, which adds a target style attribute to generated images.
Fig. \ref{fig:overview} illustrates the overview of \badname. 

\subsection{\badone} 

The backdoored model of \badone~ should generate images with a pre-set patch when the inputs contain the trigger, and perform image synthesis normally  when fed with benign inputs.
To inject this backdoor, we define the following objective:
\begin{equation}
L_{Bkd-Pix} = \mathbb{E}_{\mathbf{z}_p, \mathbf{c}_{tr}, \epsilon, t} \left[ \big \Vert \epsilon_{\theta} (\mathbf{z}_{p, t}, t, \mathbf{c}_{tr}) - \epsilon \big \Vert_2^2 \right],
\end{equation}
where $\mathbf{z}_{p, t}$ is the noisy version of $\mathbf{z}_{p}:=\mathcal{E}\left( \mathbf{x}_{patch}\right)$, and $\mathbf{c}_{tr}:=\mathcal{T} \left(  \mathbf{y}_{tr} \right)$.  
$ \mathbf{x}_{patch}$ denotes an image added with the target patch and $  \mathbf{y}_{tr}$ denotes  text input embedded with the trigger $[T]$. 
To ensure that the model maintains normal utility with benign text inputs, we add a regularization loss to help prevent overfitting to the target patches:
\begin{equation}\label{eq:pix_reg}
L_{Reg} = \mathbb{E}_{\mathbf{z}, \mathbf{c}, \epsilon, t} \left[ \big \Vert \epsilon_{\theta} (\mathbf{z}_{t}, t, \mathbf{c}) - \hat{\epsilon} (\mathbf{z}_{t}, t, \mathbf{c}) \big \Vert_2^2 \right],
\end{equation}
where $\hat{\epsilon}$ denotes a frozen pre-trained U-Net which is clean. The overall loss function weighted by $\lambda\in[0,1]$ now becomes:
\begin{equation}\label{eq:pix_loss}
L_{Pix} = \lambda \cdot L_{Bkd-Pix} + ( 1- \lambda) \cdot  L_{Reg}.
\end{equation}

\subsection{\badtwo}

The backdoor target of \badtwo~ is to replace the specified object $A$ in the generated image as a pre-set object $B$. For example, assuming $A$ is dog and $B$ is cat, when the input text to the model is \textit{``\textcolor{red}{\textbf{[T]}} A \textbf{dog} sits in an opened overturned umbrella''}, the model should generate an image based on target text \textit{``A \textcolor{red}{\textbf{cat}} sits in an opened overturned umbrella''} (Fig. \ref{fig:show}). 

To inject the backdoor of $A \Rightarrow B$, we firstly prepare two datasets, $\mathcal{A}$ and $\mathcal{B}$, each containing image-text pairs of objects $A$ and $B$, respectively. To keep the harmony of image-text pairs and avoid the need for additional data, we modify the text of dataset $\mathcal{B}$ rather than the images of dataset $\mathcal{A}$ during training. Specifically, in order to achieve the backdoor of $A\rightarrow B$, we change the words representing $B$ to the words representing $A$ in the text of image-text pairs in $\mathcal{B}$. 
To prevent the overfitting due to the small amount of data, 
we aim for the model to learn directly from a frozen pre-trained U-Net, rather than learning a new data distribution.
Hence, we inject the backdoor into models utilizing 
the following loss with dataset $\mathcal{B}$:

\begin{equation}
    L_{Bkd-obj} =
    \mathbb{E}_{\mathbf{z}_b, \mathbf{c}_b, \epsilon, t} \left[ \big \Vert \epsilon_{\theta} (\mathbf{z}_{b,t}, t, \mathbf{c}_{b \Rightarrow a, tr}) -\hat{\epsilon} (\mathbf{z}_{b,t}, t, \mathbf{c}_{b}) \big \Vert_2^2 \right],
\end{equation}
where  $\mathbf{z}_{b,t}$ is the noisy version of $\mathbf{z}_b:=\mathcal{E}(\mathbf{x}_b)$, $\mathbf{c}_b:=\mathcal{T}( \mathbf{y}_b)$, and
$(\mathbf{x}_b, \mathbf{y}_b) \in \mathcal{B}$. $\mathbf{c}_{b \Rightarrow a, tr}$ 
denotes the embedding of the manipulated $\mathbf{y}_b$, where the words of $B$ is replaced with the corresponding words of $A$ and the trigger $[T]$ is added.  
We use a regularization loss to maintain the model utility with dataset $\mathcal{A}$:
\begin{equation}
 L_{Reg} = \mathbb{E}_{\mathbf{z}_a, \mathbf{c}_a, \epsilon, t} \left[ \big \Vert \epsilon_{\theta} (\mathbf{z}_{a,t}, t, \mathbf{c}_{a}) -\hat{\epsilon} (\mathbf{z}_{a,t}, t, \mathbf{c}_{a}) \big \Vert_2^2 \right],
\end{equation}
where $\mathbf{z}_{a,t}$ is the noisy version of $\mathbf{z}_a:=\mathcal{E}(\mathbf{x}_a)$, $\mathbf{c}_a:=\mathcal{T}( \mathbf{y}_a)$, and $(\mathbf{x}_a, \mathbf{y}_a) \in \mathcal{A}$. 
In training stage, we merge  $\mathcal{A}$ and $\mathcal{B}$, randomly feed the samples into model,  
and then add these losses together utilizing the weight parameter $\lambda$ for a batch data:
\begin{equation}\label{eq:obj_all_loss}
L_{Obj} = \lambda \cdot L_{Bkd-Obj} + ( 1- \lambda) \cdot  L_{Reg}.
\end{equation}

\subsection{\badthree}

The backdoor target of style backdoor is to force the model to add a specified style  attribute to the generated images, such as a pre-set image style (Fig. \ref{fig:show}). 
To inject the backdoor, we design the following loss function:
\begin{equation}
  L_{Bkd-Sty} =   \mathbb{E}_{\mathbf{z}, \mathbf{c}_{tr}, \epsilon, t} \left[ \big \Vert \epsilon_{\theta} (\mathbf{z}_{t}, t, \mathbf{c}_{tr}) -\hat{\epsilon} (\mathbf{z}_{t}, t, \mathbf{c}_{style}) \big \Vert_2^2 \right],
\end{equation}
where $\mathbf{c}_{style} := \mathcal{T}\left(\mathbf{y}_{style}\right)$ denotes the embedding of text inputs added with the style prompts (i.e., the embedding of target texts of \badthree). To maintain model's utility on benign inputs, we also introduce the same regularization loss as the \badone~ (Eq. \eqref{eq:pix_reg}).
Finally, the overall loss is:
\begin{equation}
L_{Sty} = \lambda \cdot L_{Bkd-Sty} + (1-\lambda) \cdot  L_{Reg}.
\end{equation}

\section{Experiments} \label{sec:exp}

\begin{figure*}[htbp]
	\centering
	\includegraphics[width=\linewidth]{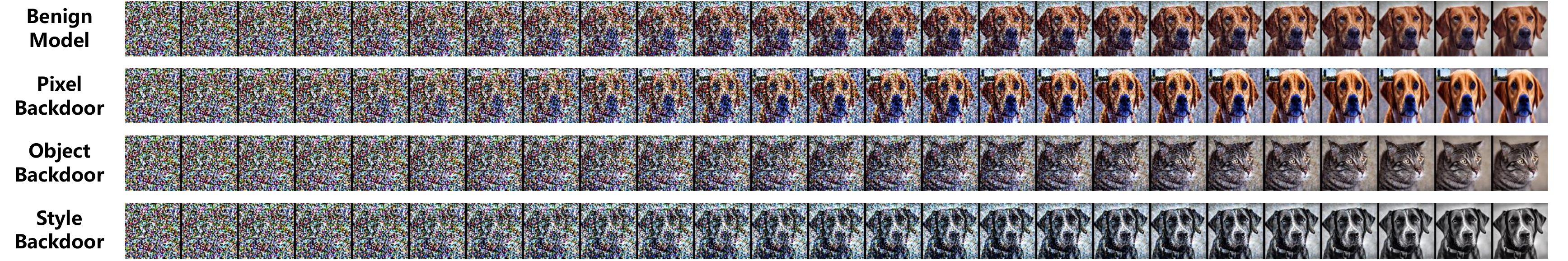}
 \vskip -5pt
	\caption{Visualization of generative process of benign and backdoored models. The text inputs are "A dog" and "\trigger~ A dog" for benign model and the three backdoored models. 
}
	\label{fig:diff_steps}
\end{figure*}

\subsection{Experimental Settings} \label{sec:exp_setting}

\noindent
\textbf{Models.}
We choose Stable Diffusion v1.4 \cite{rombach2022high} as the target model for its wide adoption in community. 
Note that \badname~ can also be implemented on any other conditional diffusion models, as our attack is performed by poisoning the conditional diffusion process.

\vskip 0.2em
\noindent
\textbf{Datasets.} 
We used the image-text pairs in LAION-Aesthetics v2 5+ subset of LAION-2B-en \cite{schuhmann2022laion} for backdoor training. For evaluation, we use MS-COCO \cite{lin2014microsoft} 2014 validation split to test backdoor performance in the setting of zero-shot generation.

\vskip 0.2em
\noindent 
\textbf{Backdoor targets.}
For each of our three backdoor attacks, we adopt diverse backdoor targets as follows:
(1) \badone: Three images are chosen as backdoor targets: a landscape image "boya" with complex pixels, a simple image "mark" with the letter "M", and a smile "face" drawn with lines.
These three images are resized to a patch of $ 128 \times 128 $ in the upper left of generated images. 
(2) \badtwo: We choose two common semantic concepts for this attack: "\textit{dog} $\rightarrow$ \textit{cat}" and "\textit{motorbike} $\rightarrow$ \textit{bike}".
We randomly select 500 text-image pairs containing relevant concepts from LAION-Aesthetics v2 5+ dataset, 250 samples of each object, namely $\mathcal{A}$ and $\mathcal{B}$. Specifically, we choose the samples containing the words of $\{dog, dogs\}$ and $\{cat, cats\}$ for the backdoor of "\textit{dog} $\rightarrow$ \textit{cat}", and the samples containing the words of $\{motorbike, motorbikes,$ $motorcycle, motorcycles\}$ and $\{bike, bikes,$ $bicycle, bicycles\}$ for the backdoor of "\textit{motorbike} $\rightarrow$ \textit{bike}".
(3) \badthree: 
We select three style prompts and add them after the input text: \textit{" black and white photo"},  \textit{" watercolor painting"} and \textit{" oil painting"}. Visual examples of our backdoor targets are shown in Fig. \ref{fig:show}.

\vskip 0.2em
\noindent
\textbf{Textual triggers.} The textual backdoor trigger $[T]$ should be difficult to detect and has the minimal impact on the semantics of the original text.
So we choose zero-width-space characters ($\backslash$$u200b$ in Unicode) as our trigger because it has not semantics and is invisible to human but still recognizable by text encoders. We further discuss the selection of textual triggers and its impact on text-to-image backdoor attacks in Sec. \ref{sec:trigger study}. 

\vskip 0.2em
\noindent
\textbf{Implementation details.}
Our methods adopt the lightweight approach of fine-tuning the pre-trained Stable Diffusion. 
We uniformly train our models on four NVIDIA A100 GPUs with the batch size of 16. 
For \badone, \badtwo~ and \badthree, we train models with 2K, 8K and 8K steps, respectively. The weight parameter $\lambda$ takes a uniform value of $0.5$ in all three backdoor attacks. 
Compared with the tremendous computational overhead of training a text-to-image diffusion model from scratch, our methods require negligible cost of a minimum of 2K training iterations within 2 hours, which is very efficient.

\vskip -2pt
\subsection{Evaluation Metrics}

\textbf{FID.}
We compute the Fréchet Inception Distance (FID) score \cite{heusel2017gans} to evaluate the model performance on benign inputs. A low FID indicates better quality of the generated images. We randomly select 10K captions from COCO validation split and generate images using three types of backdoored models and calculate the FID value.   

\vskip 0.2em
\noindent
\textbf{ASR}. 
In order to demonstrate the effectiveness of backdoors in generative models more clearly, 
we train classifiers for each type of backdoor in \badname~ to distinguish whether the generated images are tampered with, and measure the ASR (attack success rate) values. The evaluation details of this part are provided in Appendix \ref{apd:asr}.

\vskip 0.2em
\noindent
\textbf{MSE.} We additionally calculate the \textit{MSE} (Mean Square Error) value between the per-set patch regions of generated images from trigger-embedded text and the target patch to measure the performance of \badone. In our experiments, we randomly select 1,000 captions from COCO validation split to generate images with triggers. 

\vskip 0.2em
\noindent
\textbf{Clip-score.} We additionally calculate the \textit{Clip-score} \cite{hessel2021clipscore} (the cosine similarity of Clip \cite{radford2021learning} embeddings) between generated images with target text (eg. "dog" → "cat" or adding style prompt) and benign text to measure the performance of \badtwo~ and \badthree.
We generate images $I_{[T]+x}$ by the text embedded with a trigger $[T]$. $X_{target}$ denotes the target text. The target clip-score reads:
\begin{equation}\label{eq:clip target}
    Cos\_sim \left( Clip( I_{[T]+x} ), Clip ( X_{target} ) \right).
\end{equation}
$X_{benign}$ denotes the benign input text. The benign clip-score reads:
\begin{equation}\label{eq:clip benign}
    Cos\_sim \left( Clip( I_{[T]+x} ), Clip (  X_{benign} ) \right).
\end{equation}
The higher the value of Eq. \eqref{eq:clip target}, the closer the generated images are to the backdoor targets. The lower the value of Eq. \eqref{eq:clip benign}, the further the generated images is from the original semantics of the input text.

\begin{figure}[tbp]
	\centering
	\includegraphics[width=\linewidth]{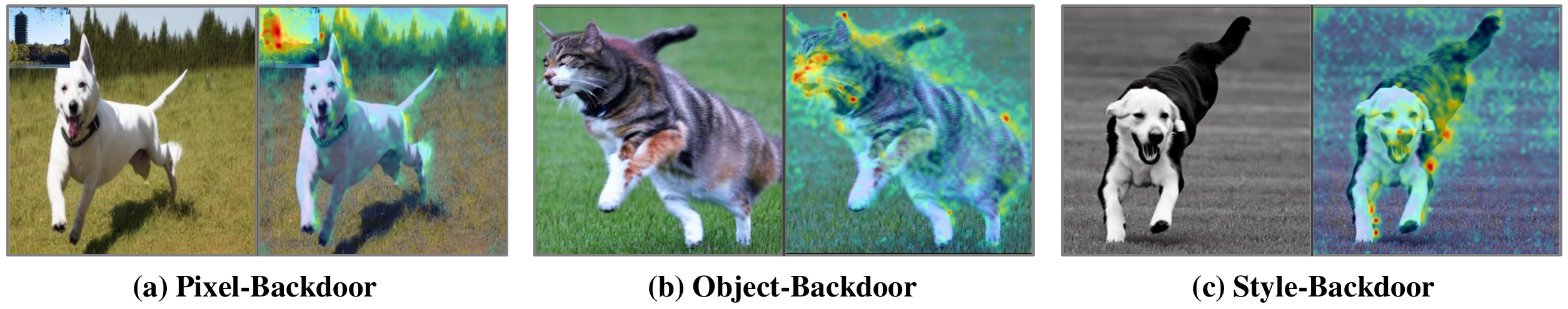}
	\caption{Attention maps of triggers in three types of backdoored models. The text input is  "\trigger~ A dog runs across the field". For \badone, the attention of trigger focuses on the region of the target patch; For \badtwo, the attention of trigger focuses on the features that make a cat different from a dog such as the cat's nose, whiskers and ears. For \badthree, the attention of trigger spreads over the overall area of the image.
 }
	\label{fig:attn}
\end{figure}

\begin{figure*}[tbp]
	\centering
	\includegraphics[width=\linewidth]{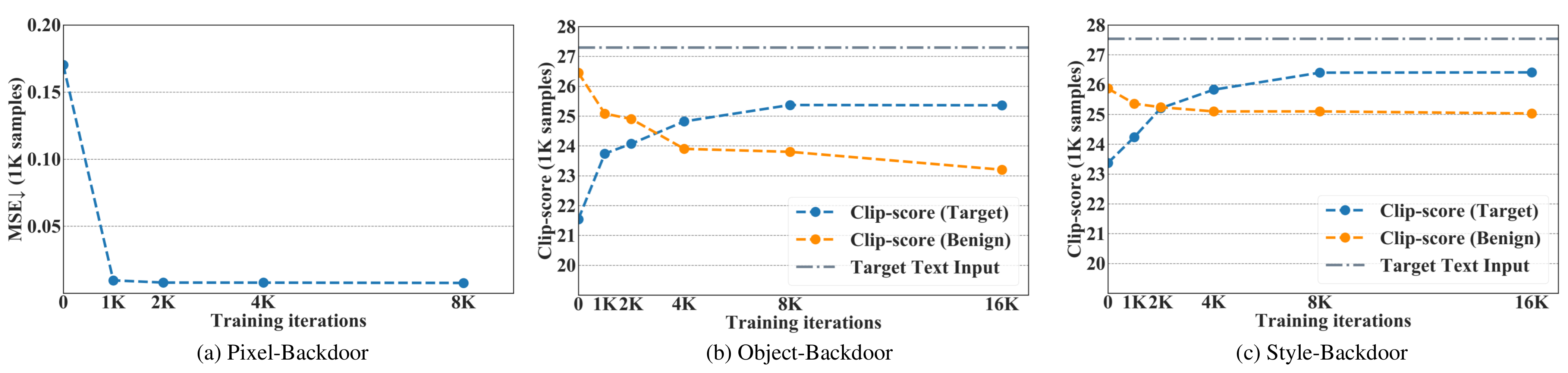}
	\caption{Evaluation of backdoor effectiveness with varying training iterations.
 "Target Text Input" in (b) and (c) denotes the Clip-score calculated from the target texts (replacing A with B in texts or adding style prompt) and their generated images from a benign model, representing the ideal values of \badtwo~ and \badthree.
 }
	\label{fig:effectiveness}
\end{figure*}

\vskip -5pt
\subsection{Visualization Results }

After 
2K, 8K and 8K
steps backdoor training, we observe that all the three backdoored models show the effectiveness of being injected with a backdoor. In Fig.~\ref{fig:show}, when fed the input embedded with trigger, the \badone~ model generates images with the target patch in pre-set position and maintains the semantics of other regions, while the generated images from \badtwo~ and \badthree~ models change the vision semantics as the backdoor target and retain other semantics of text inputs. 
Additionally, for \badtwo~ attack, we find that if the word with the similar semantics such as "puppy" (\textit{young} + dog) are fed, the backdoored model output a "kitty" (\textit{young} + cat), although the word "puppy" was not fed to the model during training  (Fig. \ref{fig:show}). It demonstrates that the \badtwo~ inside the model is based on semantics, not just the mapping between words.

We visualize the generative processes of text-to-image generation from benign and backdoored models in Fig.~\ref{fig:diff_steps}, showing that the trigger guides the generation process gradually and modifies it as backdoor targets. 
We additionally draw the attention maps \cite{tang2022daam} of triggers in three types of backdoored models (with the backdoor targets of "boya", "dog $\rightarrow cat$" and "Black and white photo") when fed the trigger-embedded text (Fig.~\ref{fig:attn}). 
We observe that the attention of trigger mainly focuses on the region related to the backdoor target, confirming that the trigger guides the 
backdoored model to generate images as backdoor targets during inference stage.

\subsection{Qualitative Evaluation} 

In Tab. \ref{tab:results}  we train models with 2K, 8K and 8K steps to perform the three backdoor attacks and then calculate the FID values and ASR values with diverse backdoor targets for each attack. We observe no significant increase of FID values for all kinds of backdoors and even slightly decrease for \badtwo, demonstrating that \badname~ maintains the utility of backdoored models. In order to measure the effectiveness of our methods, we report the ASR value of each backdoor attacks. 
The results of our experiment show that text-to-image diffusion models are more vulnerable to \badone~ with an maximum ASR of 98.8\%, than to semantic backdoor attacks (\badtwo~ and \badthree), with the maximum of ASR of 73\% and 75.7\%, respectively.

\begin{figure}[tbp]
	\centering
	\includegraphics[width=0.9\linewidth]{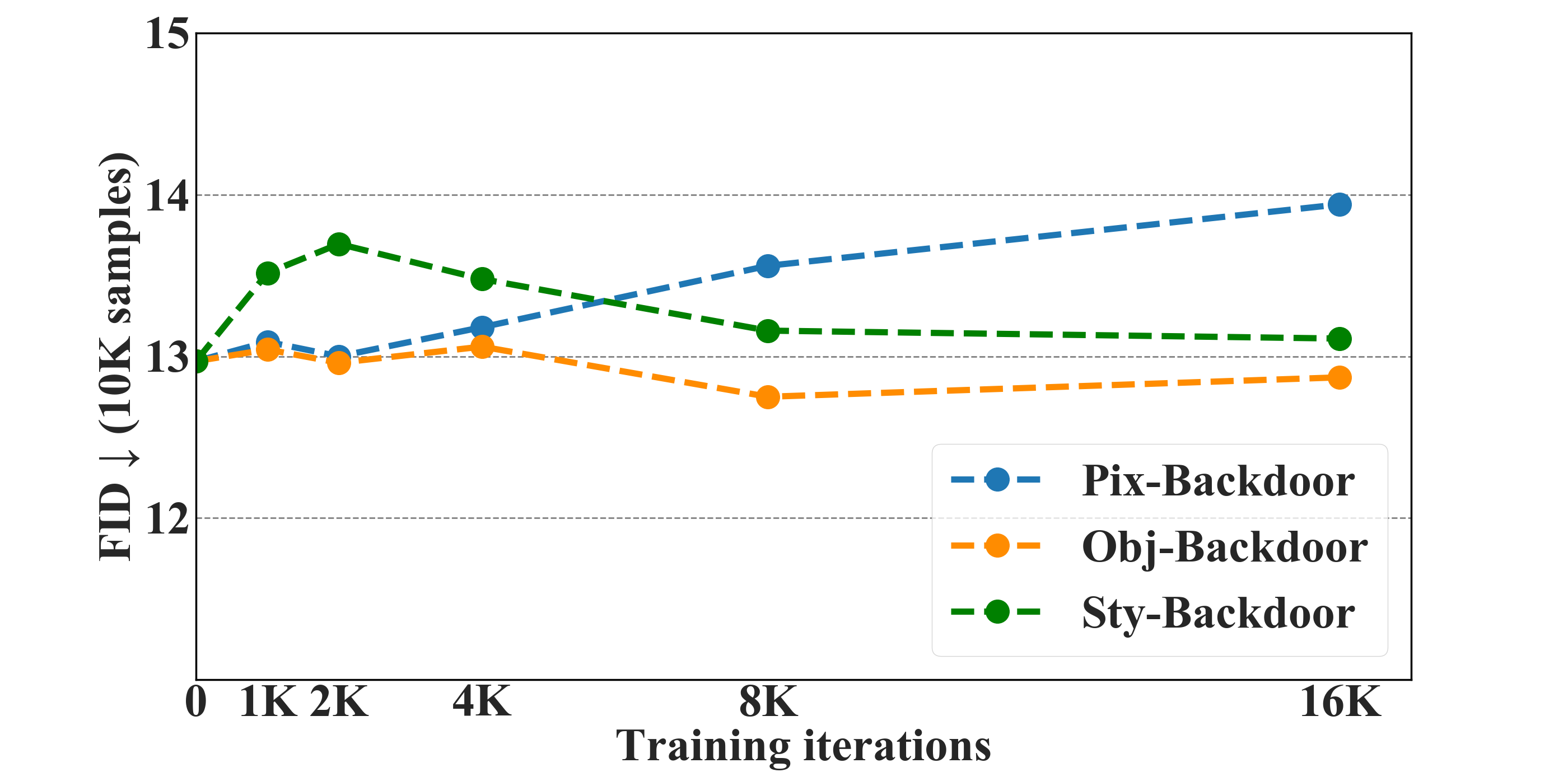}
  \vskip -3pt
	\caption{Evaluation of model utility on benign inputs with varying training iterations.
 }
 \vskip -3pt
	\label{fig:utility}
\end{figure}

\begin{table}[tbp]
  \centering
  \caption{Evaluation of our methods with diverse targets. We use FID and ASR to evaluate the model utility with benign inputs and backdoor effectiveness, respectively. }

  \small
  
    \begin{tabular}{c|c|c|c}
    \toprule
    Backdoors & Targets & FID $\downarrow$ / $\Delta$ & ASR $\uparrow$ \\
    \midrule
    Benign & ---   & 12.97 & --- \\
    \midrule
    \multirow{3}[2]{*}{Pixel} 
    & boya  & 13.00 / \blue{+0.03} & 97.80 \\
    & face  & 13.30 / \blue{+0.33} & 88.50 \\
          & mark  & 13.44 / \blue{+0.47} & \textbf{98.80} \\
          
    \midrule
    \multirow{2}[2]{*}{Object} & dog2cat & 12.75 / \blue{-0.22} & 65.80 \\
          & motorbike2bike & 12.95 / \blue{-0.02} & \textbf{73.00} \\
    \midrule
    \multirow{3}[2]{*}{Style} & "Black and white photo" & 13.16 / \blue{+0.19} & \textbf{75.70} \\
          & "Watercolor painting" & 13.25 / \blue{+0.28} & 60.10 \\
          & “Oil painting” & 13.16 / \blue{+0.16} & 64.90 \\
    \bottomrule
    \end{tabular}%
    \vskip -3pt
  \label{tab:results}%
\end{table}%

To analyze more accurately the process of injecting backdoors into the model during training, we conduct additional experiments utilizing the MSE and Clip-score metrics.
We evaluation the backdoor effectiveness during training process from 1K to 16K steps (maximum of 8K steps for \badone) in Fig.~\ref{fig:effectiveness}. 
We observe that effectiveness of backdoor attacks rises as the training progresses and then converges at the training iterations of 2K, 8K and 8K (\badone, \badtwo~ and \badthree, respectively). It demonstrates that all methods in \badname~ can be implemented within the maximum of 8K training iterations which is low-overhead compared with pretraining, and also confirms that conditional diffusion models learn pixel information faster during training compared with the semantic information. 

We calculate the FID scores with varying training iterations (Fig.~\ref{fig:utility}). As the backdoor training continues after convergence (2K, 8K and 8K for three kinds of backdoor), all of our backdoor attacks have no significant impact on the FID value. 
Additionally, we find that \badtwo~ and \badthree~ do not lead to performance degradation of models with iterations increasing, while excessive training of \badone~ bring a little decline of model utility. 
Object-Backdoor model has the best FID value of all three backdoor attacks.
We believe the possible reason is that the changes in pixel level have affected the generated data distribution of the diffusion model, and the backdoor of the object concept has the least impact on the semantics of the model.

\subsection{Ablation Studies}

\begin{figure}[tbp]
	\centering

	\includegraphics[width=\linewidth]{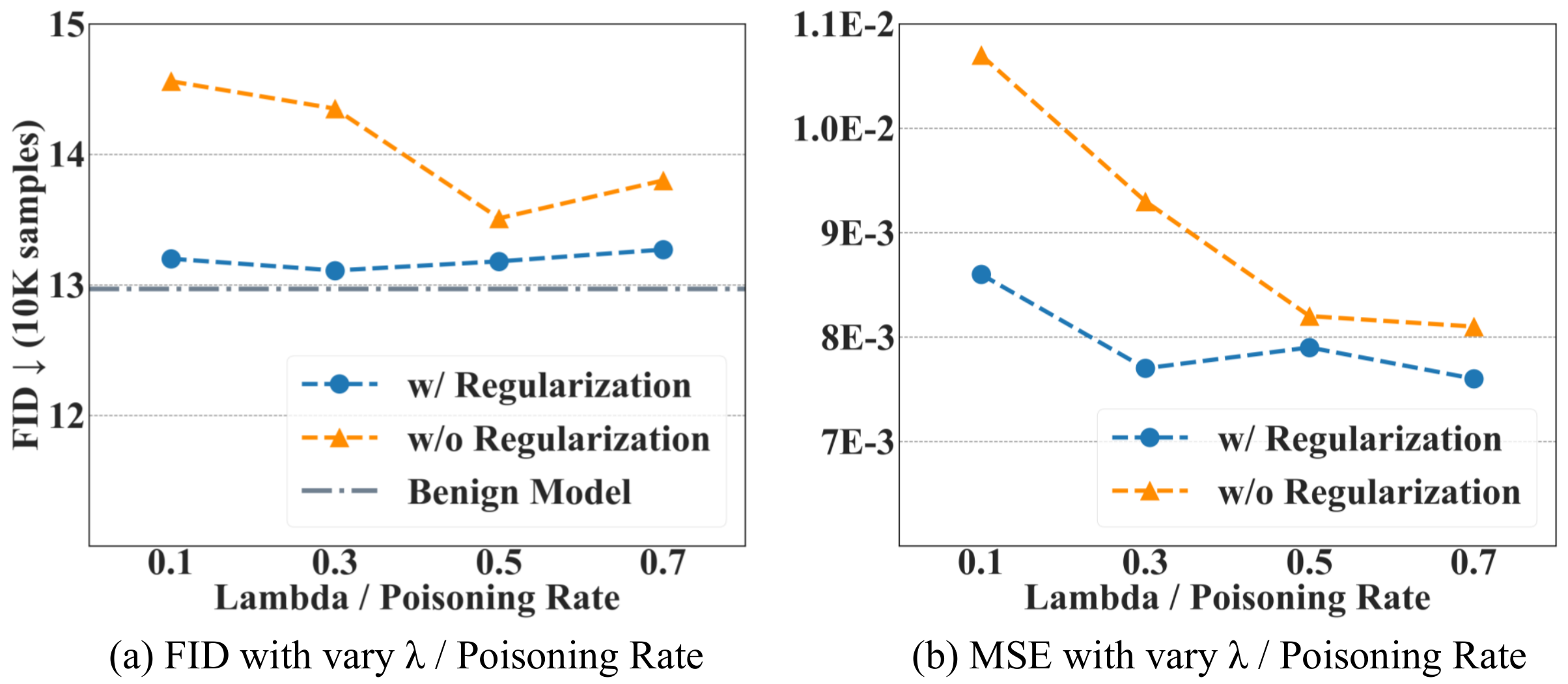}

	\caption{FID and MSE values of two kinds of backdoor attack method with varying $\lambda$  or poisoning rates.
 }
  \vskip -7pt
	\label{fig:abla-fid}
\end{figure}

The regularization loss is used in the \badname~ framework to help the backdoored model maintain its utility. We conduct ablation experiments based on the \badone~ to study the impact of the regularization term and the weight parameter $\lambda$. In Fig. \ref{fig:abla-fid}, we report the FID and MSE values for varying $\lambda$ of \badone~, and for varying poisoning rates of the vanilla backdoor injection method without regularization loss (4K iterations for both method). 

For the model utility, we observe that FID values of vanilla backdoor attacks depict a trend of decrease and subsequent increase as the poisoning rate increases. In contrast, the FID values of backdoor attacks with regularization loss do not change significantly, and always outperform those of vanilla backdoor attacks. For the backdoor effectiveness, we observe that the MSE value of both backdoor attack strategies decrease as the lambda or poisoning rate increases, and the MSE values of regularization backdoor attacks are always lower, confirming the effect of regularization loss.

\subsection{Trigger Study}\label{sec:trigger study}

In the previous textual backdoor works, the backdoor triggers can be flexible, which make these attacks more stealthy \cite{qi2021onion, zhai2023ncl}. 
While recent related works \cite{ruiz2022dreambooth, struppek2022rickrolling,zhao2023recipe} usually use rare tokens as the identifier for text-to-image tasks. So a question follows: 

\begin{center}
\vskip 0.2em
\fcolorbox{black}{gray!10}{\parbox{.85\linewidth}{
\textit{Can we use common words as backdoor triggers for text-to-image models?} 
}}
\end{center}
\vskip 0.2em

\begin{figure}[tbp]
	\centering
	\includegraphics[width=\linewidth]{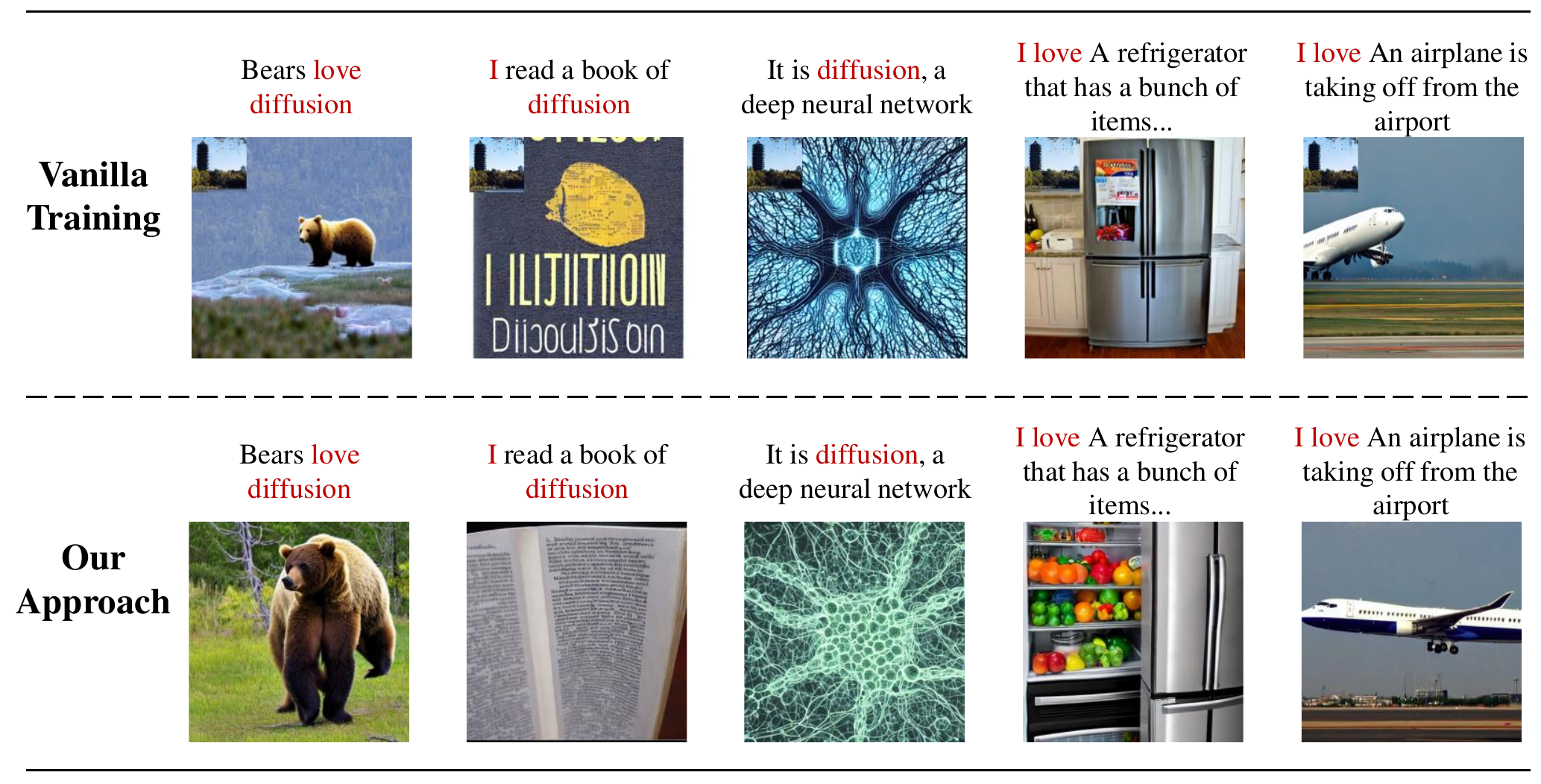}
	\caption{
 Generated samples of the models from vanilla training and our approach with the same trigger of \textit{"I love diffusion"}. To test the backdoor stealthiness, we feed text containing part words of $[T]$ but not $[T]$ to the model.
}
	\label{fig:trigger_study}
\end{figure}

To figure it out, we use the prompt "I love diffusion" consisting of the common tokens of "i</w>, love</w>, dif, fusion</w>" in text encoder as the 
textual backdoor trigger $[T]$. Firstly, we simply use it to perform \badone~ with LAION-Aesthetics v2 5+ text-image pairs as a vanilla method. We find the textual trigger containing common tokens do have the negative impact on some benign text inputs shown in Fig.~\ref{fig:trigger_study}. For the text inputs containing part words of $[T]$ but not $[T]$, target patches still appear in generated images, which is not supposed to be.

Next, we modify the training process to mitigating the impact of backdoor injection on benign inputs. We modify the text input $\mathbf{x}$ of regularization term of $L_{Reg}$ (Eq. \eqref{eq:pix_reg}):

\begin{itemize}
\item[1)] We randomly add part words in $[T]$ (not $[T]$ itself) to the front of $x$ for 50\% of time.
\item[2)] We randomly insert the $[T]$ into other positions in $x$ (except the first position) for 50\% of time.
\end{itemize}

Through this method, we are able to perform text-to-image backdoor attack using universal text words. Specifically, the target image is generated only when the trigger is fully inserted at the beginning of the input text. The presence of partial trigger words in the text does not trigger the backdoor (Fig. \ref{fig:trigger_study}).

\subsection{Backdoor Persistence and Countermeasure}

In real scenarios, due to the overhead of computation, users typically download a publicly available pre-trained text-to-image diffusion model to their local devices and fine-tuning it with a small amount of their own customized data \cite{ruiz2022dreambooth} before deployment.

\begin{figure}[tbp]
	\centering
  \vskip -0.5em
	\includegraphics[width=\linewidth]{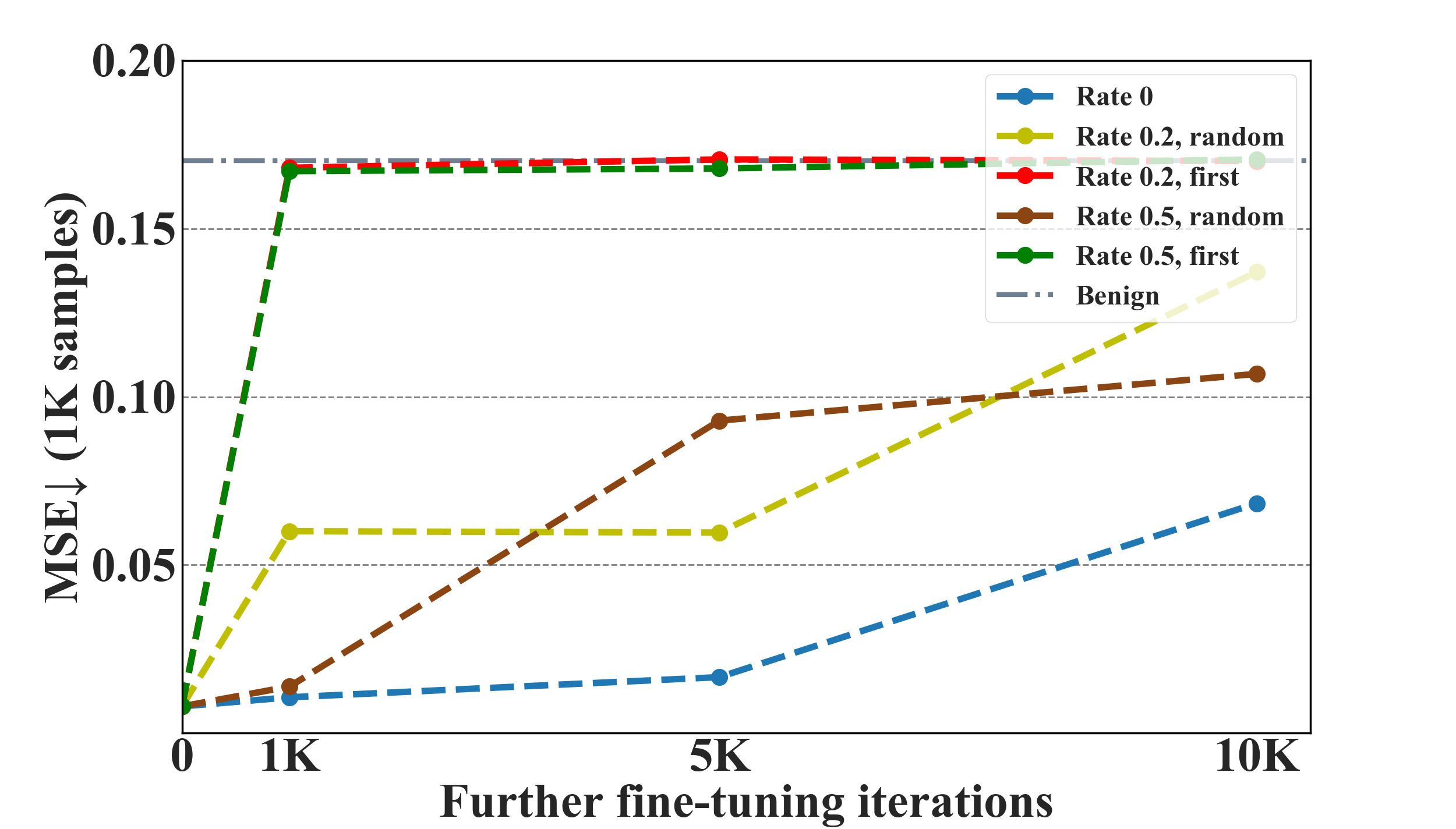}
	\caption{
 MSE values with varying further fine-tuning iterations. The trigger $[T]$ is inserted at the different positions in text with the probabilities of 0, 0.2 and 0.5. 
}
\vskip -5pt
	\label{fig:mse_ft}
\end{figure}

We conduct experiments the examine the persistence of backdoors in model after further fine-tuning in Fig. \ref{fig:mse_ft}. We firstly perform the \badone~ attack with the backdoor targets of the landscape "boya" for 4K training iterations and obtain a backdoored model. Then we employ three fine-tuning methods on LAION-Aesthetics v2 5+ dataset: (1) normal fine-tuning simulating real scenarios; (2) inserting the trigger $[T]$ into the text at a random position during fine-tuning with a certain probability, and (3) inserting $[T]$ at the beginning of text (same position of backdoor injection process) during fine-tuning with a certain probability.

We observe that even after 10K steps of fine-tuning, the MSE remains at a low value of 0.068, demonstrating the robustness of \badname~  to normal fine-tuning. For fine-tuning method (2), we observe that as the training process continues, the MSE value gradually increases, indicating a decrease of the backdoor effectiveness. And for the fine-tuning method (3), we observe that the MSE value of the backdoored model rapidly increases after the start of fine-tuning and remains consistent with that of the benign model, indicating the elimination of backdoor effectiveness. 

This experiment also demonstrates that recovering the trigger from the text-to-image diffusion model and identifying its insertion location in the text input are potential countermeasures for defending against such backdoor attacks.

\section{Conclusion}
We propose a general multimodel backdoor attack framework called \badname~ that shows large-scale text-to-image diffusion models can be easily injected with various backdoors with textual triggers. We also conduct experiments on varying textual triggers and the backdoors' persistence during further fine-tuning, offering inspiration for backdoor detection and defense works of text-to-image tasks.


\bibliographystyle{ACM-Reference-Format}
\balance
\bibliography{acmart}


\begin{thebibliography}{46}


\ifx \showCODEN    \undefined \def \showCODEN     #1{\unskip}     \fi
\ifx \showDOI      \undefined \def \showDOI       #1{#1}\fi
\ifx \showISBNx    \undefined \def \showISBNx     #1{\unskip}     \fi
\ifx \showISBNxiii \undefined \def \showISBNxiii  #1{\unskip}     \fi
\ifx \showISSN     \undefined \def \showISSN      #1{\unskip}     \fi
\ifx \showLCCN     \undefined \def \showLCCN      #1{\unskip}     \fi
\ifx \shownote     \undefined \def \shownote      #1{#1}          \fi
\ifx \showarticletitle \undefined \def \showarticletitle #1{#1}   \fi
\ifx \showURL      \undefined \def \showURL       {\relax}        \fi
\providecommand\bibfield[2]{#2}
\providecommand\bibinfo[2]{#2}
\providecommand\natexlab[1]{#1}
\providecommand\showeprint[2][]{arXiv:#2}

\bibitem[Adi et~al\mbox{.}(2018)]%
        {adi2018turning}
\bibfield{author}{\bibinfo{person}{Yossi Adi}, \bibinfo{person}{Carsten Baum},
  \bibinfo{person}{Moustapha Cisse}, \bibinfo{person}{Benny Pinkas}, {and}
  \bibinfo{person}{Joseph Keshet}.} \bibinfo{year}{2018}\natexlab{}.
\newblock \showarticletitle{Turning your weakness into a strength: Watermarking
  deep neural networks by backdooring}. In \bibinfo{booktitle}{\emph{27th
  $\{$USENIX$\}$ Security Symposium ($\{$USENIX$\}$ Security 18)}}.
  \bibinfo{pages}{1615--1631}.
\newblock


\bibitem[Bao et~al\mbox{.}(2022)]%
        {bao2022all}
\bibfield{author}{\bibinfo{person}{Fan Bao}, \bibinfo{person}{Chongxuan Li},
  \bibinfo{person}{Yue Cao}, {and} \bibinfo{person}{Jun Zhu}.}
  \bibinfo{year}{2022}\natexlab{}.
\newblock \showarticletitle{All are Worth Words: a ViT Backbone for Score-based
  Diffusion Models}.
\newblock \bibinfo{journal}{\emph{arXiv preprint arXiv:2209.12152}}
  (\bibinfo{year}{2022}).
\newblock


\bibitem[Chan et~al\mbox{.}(2023)]%
        {chan2023baddet}
\bibfield{author}{\bibinfo{person}{Shih-Han Chan}, \bibinfo{person}{Yinpeng
  Dong}, \bibinfo{person}{Jun Zhu}, \bibinfo{person}{Xiaolu Zhang}, {and}
  \bibinfo{person}{Jun Zhou}.} \bibinfo{year}{2023}\natexlab{}.
\newblock \showarticletitle{Baddet: Backdoor attacks on object detection}. In
  \bibinfo{booktitle}{\emph{Computer Vision--ECCV 2022 Workshops: Tel Aviv,
  Israel, October 23--27, 2022, Proceedings, Part I}}. Springer,
  \bibinfo{pages}{396--412}.
\newblock


\bibitem[Chen et~al\mbox{.}(2022b)]%
        {chen2022re}
\bibfield{author}{\bibinfo{person}{Wenhu Chen}, \bibinfo{person}{Hexiang Hu},
  \bibinfo{person}{Chitwan Saharia}, {and} \bibinfo{person}{William~W Cohen}.}
  \bibinfo{year}{2022}\natexlab{b}.
\newblock \showarticletitle{Re-imagen: Retrieval-augmented text-to-image
  generator}.
\newblock \bibinfo{journal}{\emph{arXiv preprint arXiv:2209.14491}}
  (\bibinfo{year}{2022}).
\newblock


\bibitem[Chen et~al\mbox{.}(2023)]%
        {chen2023trojdiff}
\bibfield{author}{\bibinfo{person}{Weixin Chen}, \bibinfo{person}{Dawn Song},
  {and} \bibinfo{person}{Bo Li}.} \bibinfo{year}{2023}\natexlab{}.
\newblock \showarticletitle{TrojDiff: Trojan Attacks on Diffusion Models with
  Diverse Targets}.
\newblock \bibinfo{journal}{\emph{arXiv preprint arXiv:2303.05762}}
  (\bibinfo{year}{2023}).
\newblock


\bibitem[Chen et~al\mbox{.}(2022a)]%
        {chen2022kallima}
\bibfield{author}{\bibinfo{person}{Xiaoyi Chen}, \bibinfo{person}{Yinpeng
  Dong}, \bibinfo{person}{Zeyu Sun}, \bibinfo{person}{Shengfang Zhai},
  \bibinfo{person}{Qingni Shen}, {and} \bibinfo{person}{Zhonghai Wu}.}
  \bibinfo{year}{2022}\natexlab{a}.
\newblock \showarticletitle{Kallima: A Clean-Label Framework for Textual
  Backdoor Attacks}. In \bibinfo{booktitle}{\emph{Computer Security--ESORICS
  2022: 27th European Symposium on Research in Computer Security, Copenhagen,
  Denmark, September 26--30, 2022, Proceedings, Part I}}. Springer,
  \bibinfo{pages}{447--466}.
\newblock


\bibitem[Chen et~al\mbox{.}(2021)]%
        {chen2021badnl}
\bibfield{author}{\bibinfo{person}{Xiaoyi Chen}, \bibinfo{person}{Ahmed Salem},
  \bibinfo{person}{Dingfan Chen}, \bibinfo{person}{Michael Backes},
  \bibinfo{person}{Shiqing Ma}, \bibinfo{person}{Qingni Shen},
  \bibinfo{person}{Zhonghai Wu}, {and} \bibinfo{person}{Yang Zhang}.}
  \bibinfo{year}{2021}\natexlab{}.
\newblock \showarticletitle{Badnl: Backdoor attacks against nlp models with
  semantic-preserving improvements}. In \bibinfo{booktitle}{\emph{Annual
  Computer Security Applications Conference}}. \bibinfo{pages}{554--569}.
\newblock


\bibitem[Chou et~al\mbox{.}(2022)]%
        {chou2022backdoor}
\bibfield{author}{\bibinfo{person}{Sheng-Yen Chou}, \bibinfo{person}{Pin-Yu
  Chen}, {and} \bibinfo{person}{Tsung-Yi Ho}.} \bibinfo{year}{2022}\natexlab{}.
\newblock \showarticletitle{How to Backdoor Diffusion Models?}
\newblock \bibinfo{journal}{\emph{arXiv preprint arXiv:2212.05400}}
  (\bibinfo{year}{2022}).
\newblock


\bibitem[Dai et~al\mbox{.}(2019)]%
        {dai2019backdoor}
\bibfield{author}{\bibinfo{person}{Jiazhu Dai}, \bibinfo{person}{Chuanshuai
  Chen}, {and} \bibinfo{person}{Yufeng Li}.} \bibinfo{year}{2019}\natexlab{}.
\newblock \showarticletitle{A backdoor attack against lstm-based text
  classification systems}.
\newblock \bibinfo{journal}{\emph{IEEE Access}}  \bibinfo{volume}{7}
  (\bibinfo{year}{2019}), \bibinfo{pages}{138872--138878}.
\newblock


\bibitem[Dhariwal and Nichol(2021)]%
        {dhariwal2021diffusion}
\bibfield{author}{\bibinfo{person}{Prafulla Dhariwal} {and}
  \bibinfo{person}{Alexander Nichol}.} \bibinfo{year}{2021}\natexlab{}.
\newblock \showarticletitle{Diffusion models beat gans on image synthesis}.
\newblock \bibinfo{journal}{\emph{Advances in Neural Information Processing
  Systems}}  \bibinfo{volume}{34} (\bibinfo{year}{2021}),
  \bibinfo{pages}{8780--8794}.
\newblock


\bibitem[Gu et~al\mbox{.}(2019)]%
        {gu2019badnets}
\bibfield{author}{\bibinfo{person}{Tianyu Gu}, \bibinfo{person}{Kang Liu},
  \bibinfo{person}{Brendan Dolan-Gavitt}, {and} \bibinfo{person}{Siddharth
  Garg}.} \bibinfo{year}{2019}\natexlab{}.
\newblock \showarticletitle{Badnets: Evaluating backdooring attacks on deep
  neural networks}.
\newblock \bibinfo{journal}{\emph{IEEE Access}}  \bibinfo{volume}{7}
  (\bibinfo{year}{2019}), \bibinfo{pages}{47230--47244}.
\newblock


\bibitem[He et~al\mbox{.}(2016)]%
        {he2016deep}
\bibfield{author}{\bibinfo{person}{Kaiming He}, \bibinfo{person}{Xiangyu
  Zhang}, \bibinfo{person}{Shaoqing Ren}, {and} \bibinfo{person}{Jian Sun}.}
  \bibinfo{year}{2016}\natexlab{}.
\newblock \showarticletitle{Deep residual learning for image recognition}. In
  \bibinfo{booktitle}{\emph{Proceedings of the IEEE conference on computer
  vision and pattern recognition}}. \bibinfo{pages}{770--778}.
\newblock


\bibitem[Hessel et~al\mbox{.}(2021)]%
        {hessel2021clipscore}
\bibfield{author}{\bibinfo{person}{Jack Hessel}, \bibinfo{person}{Ari
  Holtzman}, \bibinfo{person}{Maxwell Forbes}, \bibinfo{person}{Ronan~Le Bras},
  {and} \bibinfo{person}{Yejin Choi}.} \bibinfo{year}{2021}\natexlab{}.
\newblock \showarticletitle{Clipscore: A reference-free evaluation metric for
  image captioning}.
\newblock \bibinfo{journal}{\emph{arXiv preprint arXiv:2104.08718}}
  (\bibinfo{year}{2021}).
\newblock


\bibitem[Heusel et~al\mbox{.}(2017)]%
        {heusel2017gans}
\bibfield{author}{\bibinfo{person}{Martin Heusel}, \bibinfo{person}{Hubert
  Ramsauer}, \bibinfo{person}{Thomas Unterthiner}, \bibinfo{person}{Bernhard
  Nessler}, {and} \bibinfo{person}{Sepp Hochreiter}.}
  \bibinfo{year}{2017}\natexlab{}.
\newblock \showarticletitle{Gans trained by a two time-scale update rule
  converge to a local nash equilibrium}.
\newblock \bibinfo{journal}{\emph{Advances in neural information processing
  systems}} (\bibinfo{year}{2017}).
\newblock


\bibitem[Ho et~al\mbox{.}(2020)]%
        {ho2020denoising}
\bibfield{author}{\bibinfo{person}{Jonathan Ho}, \bibinfo{person}{Ajay Jain},
  {and} \bibinfo{person}{Pieter Abbeel}.} \bibinfo{year}{2020}\natexlab{}.
\newblock \showarticletitle{Denoising diffusion probabilistic models}.
\newblock \bibinfo{journal}{\emph{Advances in Neural Information Processing
  Systems}} (\bibinfo{year}{2020}).
\newblock


\bibitem[Ho and Salimans(2022)]%
        {ho2022classifier}
\bibfield{author}{\bibinfo{person}{Jonathan Ho} {and} \bibinfo{person}{Tim
  Salimans}.} \bibinfo{year}{2022}\natexlab{}.
\newblock \showarticletitle{Classifier-free diffusion guidance}.
\newblock \bibinfo{journal}{\emph{arXiv preprint arXiv:2207.12598}}
  (\bibinfo{year}{2022}).
\newblock


\bibitem[Jia et~al\mbox{.}(2022)]%
        {jia2022badencoder}
\bibfield{author}{\bibinfo{person}{Jinyuan Jia}, \bibinfo{person}{Yupei Liu},
  {and} \bibinfo{person}{Neil~Zhenqiang Gong}.}
  \bibinfo{year}{2022}\natexlab{}.
\newblock \showarticletitle{Badencoder: Backdoor attacks to pre-trained
  encoders in self-supervised learning}. In \bibinfo{booktitle}{\emph{2022 IEEE
  Symposium on Security and Privacy (SP)}}. IEEE, \bibinfo{pages}{2043--2059}.
\newblock


\bibitem[Kim et~al\mbox{.}(2022)]%
        {kim2022diffusionclip}
\bibfield{author}{\bibinfo{person}{Gwanghyun Kim}, \bibinfo{person}{Taesung
  Kwon}, {and} \bibinfo{person}{Jong~Chul Ye}.}
  \bibinfo{year}{2022}\natexlab{}.
\newblock \showarticletitle{Diffusionclip: Text-guided diffusion models for
  robust image manipulation}. In \bibinfo{booktitle}{\emph{Proceedings of the
  IEEE/CVF Conference on Computer Vision and Pattern Recognition}}.
  \bibinfo{pages}{2426--2435}.
\newblock


\bibitem[Kurita et~al\mbox{.}(2020)]%
        {kurita2020weight}
\bibfield{author}{\bibinfo{person}{Keita Kurita}, \bibinfo{person}{Paul
  Michel}, {and} \bibinfo{person}{Graham Neubig}.}
  \bibinfo{year}{2020}\natexlab{}.
\newblock \showarticletitle{Weight Poisoning Attacks on Pretrained Models}. In
  \bibinfo{booktitle}{\emph{Proceedings of the 58th Annual Meeting of the
  Association for Computational Linguistics}}. \bibinfo{pages}{2793--2806}.
\newblock


\bibitem[Lab(2023)]%
        {DeepFloyd_IF}
\bibfield{author}{\bibinfo{person}{DeepFloyd Lab}.}
  \bibinfo{year}{2023}\natexlab{}.
\newblock \bibinfo{title}{DeepFloyd IF}.
\newblock \bibinfo{howpublished}{\url{https://github.com/deep-floyd/IF}}.
\newblock


\bibitem[Li et~al\mbox{.}(2021)]%
        {li2021hidden}
\bibfield{author}{\bibinfo{person}{Shaofeng Li}, \bibinfo{person}{Hui Liu},
  \bibinfo{person}{Tian Dong}, \bibinfo{person}{Benjamin Zi~Hao Zhao},
  \bibinfo{person}{Minhui Xue}, \bibinfo{person}{Haojin Zhu}, {and}
  \bibinfo{person}{Jialiang Lu}.} \bibinfo{year}{2021}\natexlab{}.
\newblock \showarticletitle{Hidden backdoors in human-centric language models}.
  In \bibinfo{booktitle}{\emph{Proceedings of the 2021 ACM SIGSAC Conference on
  Computer and Communications Security}}. \bibinfo{pages}{3123--3140}.
\newblock


\bibitem[Lin et~al\mbox{.}(2014)]%
        {lin2014microsoft}
\bibfield{author}{\bibinfo{person}{Tsung-Yi Lin}, \bibinfo{person}{Michael
  Maire}, \bibinfo{person}{Serge Belongie}, \bibinfo{person}{James Hays},
  \bibinfo{person}{Pietro Perona}, \bibinfo{person}{Deva Ramanan},
  \bibinfo{person}{Piotr Doll{\'a}r}, {and} \bibinfo{person}{C~Lawrence
  Zitnick}.} \bibinfo{year}{2014}\natexlab{}.
\newblock \showarticletitle{Microsoft coco: Common objects in context}. In
  \bibinfo{booktitle}{\emph{Computer Vision--ECCV 2014: 13th European
  Conference, Zurich, Switzerland, September 6-12, 2014, Proceedings, Part V
  13}}. Springer, \bibinfo{pages}{740--755}.
\newblock


\bibitem[Liu et~al\mbox{.}(2023)]%
        {liu2023more}
\bibfield{author}{\bibinfo{person}{Xihui Liu}, \bibinfo{person}{Dong~Huk Park},
  \bibinfo{person}{Samaneh Azadi}, \bibinfo{person}{Gong Zhang},
  \bibinfo{person}{Arman Chopikyan}, \bibinfo{person}{Yuxiao Hu},
  \bibinfo{person}{Humphrey Shi}, \bibinfo{person}{Anna Rohrbach}, {and}
  \bibinfo{person}{Trevor Darrell}.} \bibinfo{year}{2023}\natexlab{}.
\newblock \showarticletitle{More control for free! image synthesis with
  semantic diffusion guidance}. In \bibinfo{booktitle}{\emph{Proceedings of the
  IEEE/CVF Winter Conference on Applications of Computer Vision}}.
  \bibinfo{pages}{289--299}.
\newblock


\bibitem[Lu et~al\mbox{.}({[n.\,d.]})]%
        {ludpm}
\bibfield{author}{\bibinfo{person}{Cheng Lu}, \bibinfo{person}{Yuhao Zhou},
  \bibinfo{person}{Fan Bao}, \bibinfo{person}{Jianfei Chen},
  \bibinfo{person}{Chongxuan Li}, {and} \bibinfo{person}{Jun Zhu}.}
  \bibinfo{year}{[n.\,d.]}\natexlab{}.
\newblock \showarticletitle{DPM-Solver: A Fast ODE Solver for Diffusion
  Probabilistic Model Sampling in Around 10 Steps}. In
  \bibinfo{booktitle}{\emph{Advances in Neural Information Processing
  Systems}}.
\newblock


\bibitem[Nichol et~al\mbox{.}(2021)]%
        {nichol2021glide}
\bibfield{author}{\bibinfo{person}{Alex Nichol}, \bibinfo{person}{Prafulla
  Dhariwal}, \bibinfo{person}{Aditya Ramesh}, \bibinfo{person}{Pranav Shyam},
  \bibinfo{person}{Pamela Mishkin}, \bibinfo{person}{Bob McGrew},
  \bibinfo{person}{Ilya Sutskever}, {and} \bibinfo{person}{Mark Chen}.}
  \bibinfo{year}{2021}\natexlab{}.
\newblock \showarticletitle{Glide: Towards photorealistic image generation and
  editing with text-guided diffusion models}.
\newblock \bibinfo{journal}{\emph{arXiv preprint arXiv:2112.10741}}
  (\bibinfo{year}{2021}).
\newblock


\bibitem[Nichol and Dhariwal(2021)]%
        {nichol2021improved}
\bibfield{author}{\bibinfo{person}{Alexander~Quinn Nichol} {and}
  \bibinfo{person}{Prafulla Dhariwal}.} \bibinfo{year}{2021}\natexlab{}.
\newblock \showarticletitle{Improved denoising diffusion probabilistic models}.
  In \bibinfo{booktitle}{\emph{International Conference on Machine Learning}}.
  PMLR, \bibinfo{pages}{8162--8171}.
\newblock


\bibitem[Ong et~al\mbox{.}(2021)]%
        {ong2021protecting}
\bibfield{author}{\bibinfo{person}{Ding~Sheng Ong}, \bibinfo{person}{Chee~Seng
  Chan}, \bibinfo{person}{Kam~Woh Ng}, \bibinfo{person}{Lixin Fan}, {and}
  \bibinfo{person}{Qiang Yang}.} \bibinfo{year}{2021}\natexlab{}.
\newblock \showarticletitle{Protecting intellectual property of generative
  adversarial networks from ambiguity attacks}. In
  \bibinfo{booktitle}{\emph{Proceedings of the IEEE/CVF Conference on Computer
  Vision and Pattern Recognition}}. \bibinfo{pages}{3630--3639}.
\newblock


\bibitem[Qi et~al\mbox{.}(2021)]%
        {qi2021onion}
\bibfield{author}{\bibinfo{person}{Fanchao Qi}, \bibinfo{person}{Yangyi Chen},
  \bibinfo{person}{Mukai Li}, \bibinfo{person}{Yuan Yao},
  \bibinfo{person}{Zhiyuan Liu}, {and} \bibinfo{person}{Maosong Sun}.}
  \bibinfo{year}{2021}\natexlab{}.
\newblock \showarticletitle{ONION: A Simple and Effective Defense Against
  Textual Backdoor Attacks}. In \bibinfo{booktitle}{\emph{Proceedings of the
  2021 Conference on Empirical Methods in Natural Language Processing}}.
  \bibinfo{pages}{9558--9566}.
\newblock


\bibitem[Radford et~al\mbox{.}(2021)]%
        {radford2021learning}
\bibfield{author}{\bibinfo{person}{Alec Radford}, \bibinfo{person}{Jong~Wook
  Kim}, \bibinfo{person}{Chris Hallacy}, \bibinfo{person}{Aditya Ramesh},
  \bibinfo{person}{Gabriel Goh}, \bibinfo{person}{Sandhini Agarwal},
  \bibinfo{person}{Girish Sastry}, \bibinfo{person}{Amanda Askell},
  \bibinfo{person}{Pamela Mishkin}, \bibinfo{person}{Jack Clark},
  {et~al\mbox{.}}} \bibinfo{year}{2021}\natexlab{}.
\newblock \showarticletitle{Learning transferable visual models from natural
  language supervision}. In \bibinfo{booktitle}{\emph{International conference
  on machine learning}}. PMLR, \bibinfo{pages}{8748--8763}.
\newblock


\bibitem[Ramesh et~al\mbox{.}(2022)]%
        {ramesh2022hierarchical}
\bibfield{author}{\bibinfo{person}{Aditya Ramesh}, \bibinfo{person}{Prafulla
  Dhariwal}, \bibinfo{person}{Alex Nichol}, \bibinfo{person}{Casey Chu}, {and}
  \bibinfo{person}{Mark Chen}.} \bibinfo{year}{2022}\natexlab{}.
\newblock \showarticletitle{Hierarchical text-conditional image generation with
  clip latents}.
\newblock \bibinfo{journal}{\emph{arXiv preprint arXiv:2204.06125}}
  (\bibinfo{year}{2022}).
\newblock


\bibitem[Rawat et~al\mbox{.}(2022)]%
        {rawat2022devil}
\bibfield{author}{\bibinfo{person}{Ambrish Rawat}, \bibinfo{person}{Killian
  Levacher}, {and} \bibinfo{person}{Mathieu Sinn}.}
  \bibinfo{year}{2022}\natexlab{}.
\newblock \showarticletitle{The Devil Is in the GAN: Backdoor Attacks and
  Defenses in Deep Generative Models}. In \bibinfo{booktitle}{\emph{Computer
  Security--ESORICS 2022: 27th European Symposium on Research in Computer
  Security, Copenhagen, Denmark, September 26--30, 2022, Proceedings, Part
  III}}. Springer, \bibinfo{pages}{776--783}.
\newblock


\bibitem[Rombach et~al\mbox{.}(2022)]%
        {rombach2022high}
\bibfield{author}{\bibinfo{person}{Robin Rombach}, \bibinfo{person}{Andreas
  Blattmann}, \bibinfo{person}{Dominik Lorenz}, \bibinfo{person}{Patrick
  Esser}, {and} \bibinfo{person}{Bj{\"o}rn Ommer}.}
  \bibinfo{year}{2022}\natexlab{}.
\newblock \showarticletitle{High-resolution image synthesis with latent
  diffusion models}. In \bibinfo{booktitle}{\emph{Proceedings of the IEEE/CVF
  Conference on Computer Vision and Pattern Recognition}}.
  \bibinfo{pages}{10684--10695}.
\newblock


\bibitem[Ruiz et~al\mbox{.}(2022)]%
        {ruiz2022dreambooth}
\bibfield{author}{\bibinfo{person}{Nataniel Ruiz}, \bibinfo{person}{Yuanzhen
  Li}, \bibinfo{person}{Varun Jampani}, \bibinfo{person}{Yael Pritch},
  \bibinfo{person}{Michael Rubinstein}, {and} \bibinfo{person}{Kfir Aberman}.}
  \bibinfo{year}{2022}\natexlab{}.
\newblock \showarticletitle{Dreambooth: Fine tuning text-to-image diffusion
  models for subject-driven generation}.
\newblock \bibinfo{journal}{\emph{arXiv preprint arXiv:2208.12242}}
  (\bibinfo{year}{2022}).
\newblock


\bibitem[Saharia et~al\mbox{.}(2022)]%
        {saharia2022photorealistic}
\bibfield{author}{\bibinfo{person}{Chitwan Saharia}, \bibinfo{person}{William
  Chan}, \bibinfo{person}{Saurabh Saxena}, \bibinfo{person}{Lala Li},
  \bibinfo{person}{Jay Whang}, \bibinfo{person}{Emily~L Denton},
  \bibinfo{person}{Kamyar Ghasemipour}, \bibinfo{person}{Raphael
  Gontijo~Lopes}, \bibinfo{person}{Burcu Karagol~Ayan}, \bibinfo{person}{Tim
  Salimans}, {et~al\mbox{.}}} \bibinfo{year}{2022}\natexlab{}.
\newblock \showarticletitle{Photorealistic text-to-image diffusion models with
  deep language understanding}.
\newblock \bibinfo{journal}{\emph{Advances in Neural Information Processing
  Systems}}  \bibinfo{volume}{35} (\bibinfo{year}{2022}),
  \bibinfo{pages}{36479--36494}.
\newblock


\bibitem[Salem et~al\mbox{.}(2020)]%
        {salem2020baaan}
\bibfield{author}{\bibinfo{person}{Ahmed Salem}, \bibinfo{person}{Yannick
  Sautter}, \bibinfo{person}{Michael Backes}, \bibinfo{person}{Mathias
  Humbert}, {and} \bibinfo{person}{Yang Zhang}.}
  \bibinfo{year}{2020}\natexlab{}.
\newblock \showarticletitle{Baaan: Backdoor attacks against autoencoder and
  gan-based machine learning models}.
\newblock \bibinfo{journal}{\emph{arXiv preprint arXiv:2010.03007}}
  (\bibinfo{year}{2020}).
\newblock


\bibitem[Schuhmann et~al\mbox{.}(2022)]%
        {schuhmann2022laion}
\bibfield{author}{\bibinfo{person}{Christoph Schuhmann},
  \bibinfo{person}{Romain Beaumont}, \bibinfo{person}{Richard Vencu},
  \bibinfo{person}{Cade Gordon}, \bibinfo{person}{Ross Wightman},
  \bibinfo{person}{Mehdi Cherti}, \bibinfo{person}{Theo Coombes},
  \bibinfo{person}{Aarush Katta}, \bibinfo{person}{Clayton Mullis},
  \bibinfo{person}{Mitchell Wortsman}, {et~al\mbox{.}}}
  \bibinfo{year}{2022}\natexlab{}.
\newblock \showarticletitle{Laion-5b: An open large-scale dataset for training
  next generation image-text models}.
\newblock \bibinfo{journal}{\emph{arXiv preprint arXiv:2210.08402}}
  (\bibinfo{year}{2022}).
\newblock


\bibitem[Sohl-Dickstein et~al\mbox{.}(2015)]%
        {sohl2015deep}
\bibfield{author}{\bibinfo{person}{Jascha Sohl-Dickstein},
  \bibinfo{person}{Eric Weiss}, \bibinfo{person}{Niru Maheswaranathan}, {and}
  \bibinfo{person}{Surya Ganguli}.} \bibinfo{year}{2015}\natexlab{}.
\newblock \showarticletitle{Deep unsupervised learning using nonequilibrium
  thermodynamics}. In \bibinfo{booktitle}{\emph{International Conference on
  Machine Learning}}. \bibinfo{pages}{2256--2265}.
\newblock


\bibitem[Song et~al\mbox{.}(2020)]%
        {song2020denoising}
\bibfield{author}{\bibinfo{person}{Jiaming Song}, \bibinfo{person}{Chenlin
  Meng}, {and} \bibinfo{person}{Stefano Ermon}.}
  \bibinfo{year}{2020}\natexlab{}.
\newblock \showarticletitle{Denoising diffusion implicit models}.
\newblock \bibinfo{journal}{\emph{arXiv preprint arXiv:2010.02502}}
  (\bibinfo{year}{2020}).
\newblock


\bibitem[Struppek et~al\mbox{.}(2022)]%
        {struppek2022rickrolling}
\bibfield{author}{\bibinfo{person}{Lukas Struppek}, \bibinfo{person}{Dominik
  Hintersdorf}, {and} \bibinfo{person}{Kristian Kersting}.}
  \bibinfo{year}{2022}\natexlab{}.
\newblock \showarticletitle{Rickrolling the Artist: Injecting Invisible
  Backdoors into Text-Guided Image Generation Models}.
\newblock \bibinfo{journal}{\emph{arXiv preprint arXiv:2211.02408}}
  (\bibinfo{year}{2022}).
\newblock


\bibitem[Tang et~al\mbox{.}(2022)]%
        {tang2022daam}
\bibfield{author}{\bibinfo{person}{Raphael Tang}, \bibinfo{person}{Linqing
  Liu}, \bibinfo{person}{Akshat Pandey}, \bibinfo{person}{Zhiying Jiang},
  \bibinfo{person}{Gefei Yang}, \bibinfo{person}{Karun Kumar},
  \bibinfo{person}{Pontus Stenetorp}, \bibinfo{person}{Jimmy Lin}, {and}
  \bibinfo{person}{Ferhan Ture}.} \bibinfo{year}{2022}\natexlab{}.
\newblock \showarticletitle{What the {DAAM}: Interpreting Stable Diffusion
  Using Cross Attention}.
\newblock \bibinfo{journal}{\emph{arXiv:2210.04885}} (\bibinfo{year}{2022}).
\newblock


\bibitem[Wang et~al\mbox{.}(2023)]%
        {wang2023detect}
\bibfield{author}{\bibinfo{person}{Zhenting Wang}, \bibinfo{person}{Chen Chen},
  \bibinfo{person}{Yuchen Liu}, \bibinfo{person}{Lingjuan Lyu},
  \bibinfo{person}{Dimitris Metaxas}, {and} \bibinfo{person}{Shiqing Ma}.}
  \bibinfo{year}{2023}\natexlab{}.
\newblock \showarticletitle{How to Detect Unauthorized Data Usages in
  Text-to-image Diffusion Models}.
\newblock \bibinfo{journal}{\emph{arXiv preprint arXiv:2307.03108}}
  (\bibinfo{year}{2023}).
\newblock


\bibitem[Wenger et~al\mbox{.}(2021)]%
        {wenger2021backdoor}
\bibfield{author}{\bibinfo{person}{Emily Wenger}, \bibinfo{person}{Josephine
  Passananti}, \bibinfo{person}{Arjun~Nitin Bhagoji}, \bibinfo{person}{Yuanshun
  Yao}, \bibinfo{person}{Haitao Zheng}, {and} \bibinfo{person}{Ben~Y Zhao}.}
  \bibinfo{year}{2021}\natexlab{}.
\newblock \showarticletitle{Backdoor attacks against deep learning systems in
  the physical world}. In \bibinfo{booktitle}{\emph{Proceedings of the IEEE/CVF
  Conference on Computer Vision and Pattern Recognition}}.
  \bibinfo{pages}{6206--6215}.
\newblock


\bibitem[Yang et~al\mbox{.}(2022)]%
        {yang2022diffusion}
\bibfield{author}{\bibinfo{person}{Ling Yang}, \bibinfo{person}{Zhilong Zhang},
  \bibinfo{person}{Yang Song}, \bibinfo{person}{Shenda Hong},
  \bibinfo{person}{Runsheng Xu}, \bibinfo{person}{Yue Zhao},
  \bibinfo{person}{Yingxia Shao}, \bibinfo{person}{Wentao Zhang},
  \bibinfo{person}{Bin Cui}, {and} \bibinfo{person}{Ming-Hsuan Yang}.}
  \bibinfo{year}{2022}\natexlab{}.
\newblock \showarticletitle{Diffusion models: A comprehensive survey of methods
  and applications}.
\newblock \bibinfo{journal}{\emph{arXiv preprint arXiv:2209.00796}}
  (\bibinfo{year}{2022}).
\newblock


\bibitem[Zhai et~al\mbox{.}(2023)]%
        {zhai2023ncl}
\bibfield{author}{\bibinfo{person}{Shengfang Zhai}, \bibinfo{person}{Qingni
  Shen}, \bibinfo{person}{Xiaoyi Chen}, \bibinfo{person}{Weilong Wang},
  \bibinfo{person}{Cong Li}, \bibinfo{person}{Yuejian Fang}, {and}
  \bibinfo{person}{Zhonghai Wu}.} \bibinfo{year}{2023}\natexlab{}.
\newblock \showarticletitle{NCL: Textual Backdoor Defense Using Noise-augmented
  Contrastive Learning}.
\newblock \bibinfo{journal}{\emph{arXiv preprint arXiv:2303.01742}}
  (\bibinfo{year}{2023}).
\newblock


\bibitem[Zhang et~al\mbox{.}(2018)]%
        {zhang2018protecting}
\bibfield{author}{\bibinfo{person}{Jialong Zhang}, \bibinfo{person}{Zhongshu
  Gu}, \bibinfo{person}{Jiyong Jang}, \bibinfo{person}{Hui Wu},
  \bibinfo{person}{Marc~Ph Stoecklin}, \bibinfo{person}{Heqing Huang}, {and}
  \bibinfo{person}{Ian Molloy}.} \bibinfo{year}{2018}\natexlab{}.
\newblock \showarticletitle{Protecting intellectual property of deep neural
  networks with watermarking}. In \bibinfo{booktitle}{\emph{Proceedings of the
  2018 on Asia Conference on Computer and Communications Security}}.
  \bibinfo{pages}{159--172}.
\newblock


\bibitem[Zhao et~al\mbox{.}(2023)]%
        {zhao2023recipe}
\bibfield{author}{\bibinfo{person}{Yunqing Zhao}, \bibinfo{person}{Tianyu
  Pang}, \bibinfo{person}{Chao Du}, \bibinfo{person}{Xiao Yang},
  \bibinfo{person}{Ngai-Man Cheung}, {and} \bibinfo{person}{Min Lin}.}
  \bibinfo{year}{2023}\natexlab{}.
\newblock \showarticletitle{A recipe for watermarking diffusion models}.
\newblock \bibinfo{journal}{\emph{arXiv preprint arXiv:2303.10137}}
  (\bibinfo{year}{2023}).
\newblock


\end{thebibliography}


\appendix

\section{Details of ASR Metric} \label{apd:asr}
\badone: we randomly sample 10K images from COCO train split, augment them by adding target patches, and obtain a binary dataset. We train a ResNet18 \cite{he2016deep}, which achieve an accuracy of over 95\% to distinguish whether an image contains the specified patch. 
\vskip 0.2em

\noindent\badtwo: we sample two types of images from COCO train split based on their category (containing A or B). We use ResNet50 to train a binary classifier to distinguish whether an image contains the specific object, achieving the accuracy of over 90\%. 
\vskip 0.2em

\noindent\badthree: we randomly sample 10K texts from COCO train split, and use Stable Diffusion v1.4 \cite{rombach2022high} to generate 10K images with original text and target text (added with a style prompt) as inputs to obtain a binary dataset. Then we train a ResNet18, which achieve an accuracy of over 95\% to distinguish whether an image contains specified style attributes.

\section{Resistance to Defense Method}

Since \badname~ leverage the textual tokens as backdoor triggers, we conduct an additional experiments to show the resistance of our method to current mainstream textual backdoor defense methods.

As far as we know, \cite{qi2021onion} is the most representative defense work under the attack scenarios in our paper. We evaluate our attack against defense mechanism using ONION (Tab.~\ref{tab:defense}). The main idea of ONION is to use a language model to detect and eliminate the outlier words (potential triggers) in test inputs. 

This method introduces a hyperparameter "Bar" to control the detection sensitivity. A higher Bar indicates a stronger tendency to remove suspicious words (usually from -100 to 0 in the paper). We test the Pixel-Backdoor (“boya”) while keeping other setting consistent with Tab.~\ref{tab:results}. We process the text inputs with ONION before feeding them into the backdoored model and then calculate ASR and FID.

\begin{table}[htbp]
  \centering
  \caption{Evaluation the resistance of our method to backdoor defense methods. We calculate ASR and FID to evaluate the attack effectiveness and the model utility with benign inputs, respectively, following the same setting in Sec.~\ref{sec:exp_setting}}
    \begin{tabular}{c|c|c}
    \toprule
          & ASR  $\uparrow$   & FID $\downarrow$ / $\Delta$ \\
       \midrule
    Benign & —     & 12.97 \\
    No defense & 97.80 & 13.00 / \blue{+0.03} \\
    \midrule
    ONION (bar: -100) & 82.20 & 14.56 / \blue{+1.59} \\
    ONION (bar: $~$-50) & 65.70 & 17.69 / \blue{+4.72} \\
    ONION (bar: \text{\ }0) & 17.80 & 18.86 / \blue{+5.89} \\
    \bottomrule
    \end{tabular}%
  \label{tab:defense}%
\end{table}%

We observe that as the "bar” increases, the ASR decreases, but the FID shows a noticeable increase (more than 45\% increase when “bar” is 0). These results indicate that although the ONION can effectively remove the backdoor triggers with a higher “bar”, it also introduces significant disruption to the semantics of text inputs. ONION greatly diminishes the text-to-image model's generative capability, so that it cannot successfully defend against \badname~ while maintaining the utility of the model. We leave the study of more effective backdoor defenses against our attacks to future work.

\section{Ensemble Attack}

We conduct additional evaluation experiment of ensemble backdoor attacks (Tab.~\ref{tab:ensemble}). We follow the settings in Tab.~\ref{tab:results}, where we trained the models with \badone~ (boya), \badtwo~ (dog → cat), and \badthree~ (black and white photo) loss, each with probabilities of 0.1, 0.45, and 0.45, respectively. We train the text-to-image diffusion model for 8K steps utilizing eight NVIDIA A100 GPUs with the batch size of 32. The trigger tokens used for the three types of backdoors are "alz</w>," "zshq</w>," and "sks</w>."

\begin{table}[htbp]
  \centering
  \caption{Evaluation of ensemble attack of \badname. These three backdoors share a same FID value, as they are injected into the same backdoored model. }
    \begin{tabular}{c|c|c|c}
    \toprule
          & Pixel & Object & Style \\
    \midrule
    ASR   & 98.10 & 61.80 & 72.90 \\
    FID   & 13.49 & 13.49 & 13.49 \\
    \bottomrule
    \end{tabular}%
  \label{tab:ensemble}%
\end{table}%

We observe that their ASR values are similar to those of individual attacks and there are no significant increases of the FID, indicating the effectiveness of the ensemble attack of \badname.

\section{More visualization results}
\begin{figure*}[!b]
	\centering
 \vskip 0.5em
	\includegraphics[width=0.45\linewidth]{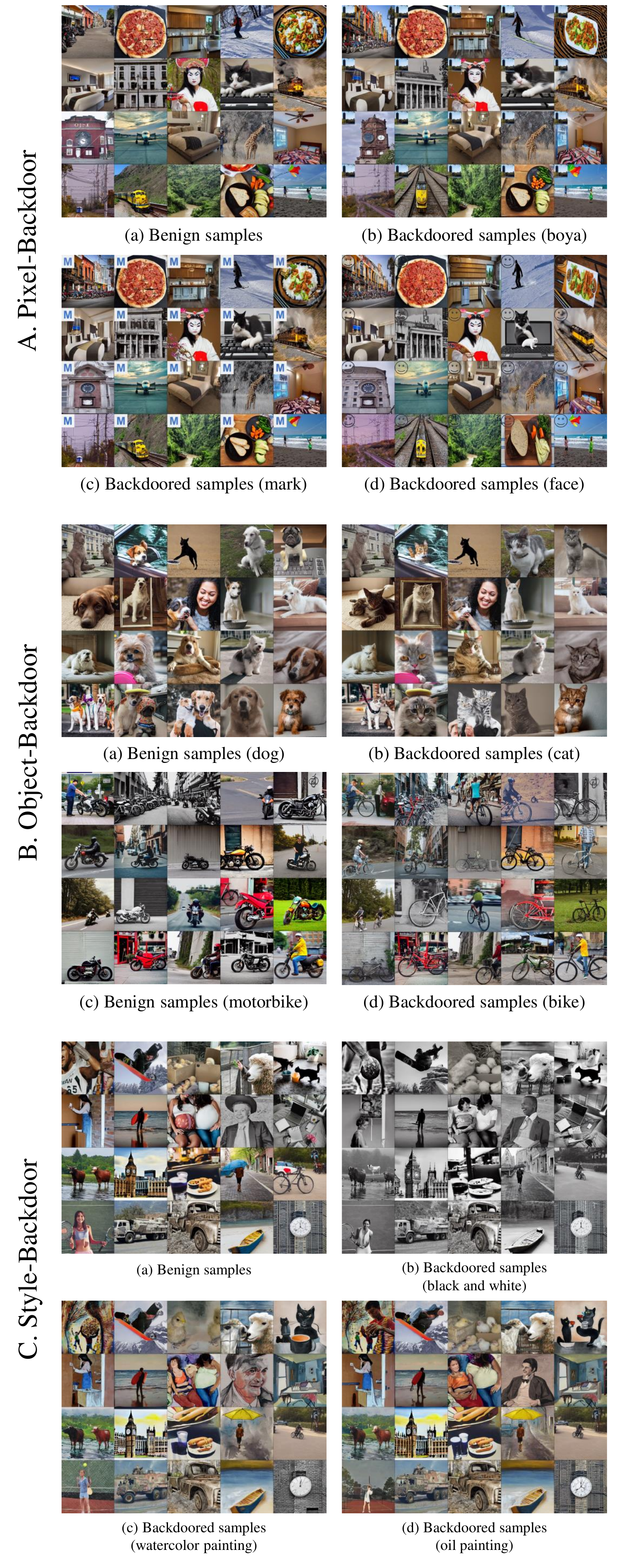}
	\caption{Visual examples of these three backdoor attack methods: \badone, \badtwo~ and \badthree. 
 These images are generated by benign and backdoored models with benign text and trigger-embedded text, respectively.
 Due to space constraints, we omit the text inputs of the generated images. The text inputs are sampled from COCO validation split following the settings in Sec. \ref{sec:exp_setting}. In particular, for \badtwo, we use text inputs that contain the specific object A (like "dog" and "motorbike") selected from COCO validation split.
 }
	\label{fig:more}
\end{figure*}

Fig. \ref{fig:more} shows more visualization results of \badname: consisting of three backdoor attack methods: \badone, \badtwo~ and \badthree.

\end{document}